\documentclass[journal]{IEEEtran}

\usepackage{amsfonts,amsmath,epsfig,subfigure,colordvi,fancybox,
balance,amssymb,graphicx,bbold}
\usepackage{amssymb,mathrsfs}
\usepackage{graphicx}
\usepackage{subfigure}
\usepackage{multirow}
\usepackage{cite}
\usepackage{color}
\usepackage[labelsep=space]{caption}
\usepackage{algorithmicx,algorithm}
\usepackage{amssymb,mathrsfs}
\usepackage{enumerate}
\usepackage{setspace}
\usepackage{bm}
\usepackage{caption}
\usepackage{graphicx}
\usepackage{epstopdf}
\usepackage{amsmath}
\usepackage{array}
\usepackage{fancyhdr}
\usepackage{cite}
\usepackage{stfloats}

\newtheorem{thm}{Theorem}[section]

\newtheorem{lem}{Lemma}[section]
\newtheorem{cor}{Corollary}[section]
\newtheorem{rem}{Remark}[section]
\newtheorem{defn}{Definition}[section]

\newtheorem{asm}{Assumption}[section]

\newcommand{\beq}[1]{\begin{equation} \label{#1}}
\newcommand{\eeq}{\end{equation}}
\newcommand{\bed}{\begin{displaymath}}
\newcommand{\eed}{\end{displaymath}}
\newcommand{\bea}{\bed\begin{array}{rl}}
\newcommand{\eea}{\end{array}\eed}

\newcommand{\barray}{\begin{array}{ll}}
\newcommand{\earray}{\end{array}}
\newcommand{\bedd}{\bed\begin{array}{l}}
\newcommand{\eedd}{\end{array}\eed}

\def\({\left(}
\def\){\right)}

\begin{document}

\title{Stability analysis of distributed Kalman filtering
	algorithm for stochastic regression model}

\author{Siyu Xie,~\IEEEmembership{}
          Die Gan,~\IEEEmembership{} Zhixin Liu~\IEEEmembership{}
\thanks{S. Xie is with School of Aeronautics and Astronautic, University of Electronic Science
	and Technology of China, Chengdu, 611731. Email: \emph{syxie@uestc.edu.cn}

D. Gan is with College of Artificial Intelligence, Nankai University, Tianjin, 300350,
China.  Email: \emph{gandie@amss.ac.cn} (Corresponding author)

Z. Liu is with Key Laboratory of Systems and Control, Academy of Mathematics and Systems Science, 100190, Beijing, China. \emph{lzx@amss.ac.cn}
}
}


\maketitle

\begin{abstract}
In this paper, a distributed Kalman filtering (DKF) algorithm is proposed based on a diffusion strategy, which is used to track an unknown signal process in sensor networks cooperatively. Unlike the centralized algorithms, no fusion center is need here, which implies that the DKF algorithm is more robust and scalable. Moreover, the stability of the DKF algorithm is established under non-independent and non-stationary signal conditions. The cooperative information condition used in the paper shows that even if any sensor cannot track the unknown signal individually, the DKF algorithm can be utilized to fulfill the estimation task in a cooperative way. Finally, we illustrate the cooperative property of the DKF algorithm by using a simulation example.
\end{abstract}

\begin{IEEEkeywords}
Kalman filtering algorithm, distributed adaptive filters, $L_p$-exponentially stability, cooperative information condition
\end{IEEEkeywords}

\IEEEpeerreviewmaketitle

\section{Introduction}

Nowadays, more and more data can be collected through sensor networks, and estimating or tracking an unknown signal process of interest based on the collected data has attracted a lot of research attention. Basically, there are two different ways to  process the data, i.e., the centralized and distributed method. For the centralized processing method, measurements or estimates from all sensors over the network need to be transferred to a  fusion center, which may not be feasible due to limited communication capabilities, energy consumptions, packet losses or privacy considerations. 
Moreover, this method lacks robustness, since whenever the fusion center fails the whole network collapses. Because of these drawbacks, the distributed processing approach arises, where each sensor utilizes the local observations and the information derived from its neighbors to estimate the unknown parameters, which is more robust and scalable compared with the centralized case. Moreover, distributed estimation algorithms may achieve the same performance with the centralized case by optimizing the adjacency matrix.

Note that different kinds of distributed estimation algorithms can be obtained by combining different cooperative strategies and different estimation algorithms. For examples, incremental LMS \cite{Sayed2014Adaptive, Khalili2017Performance}, consensus LMS \cite{Xie2018Auto, Xie2018SIAM}, diffusion LMS \cite{Xie2018TAC, Piggott2016Diffusion, Nosrati2015Adaptive, Tu2012Diffusion, Vah2018AnalysisLMS, Har2019On}, incremental LS \cite{Sayed2006Distributed, Chaves2013Distributed}, consensus LS \cite{Mateos2012Distributed, Mateos2009Distributed, Gratton2019Consensus}, diffusion LS \cite{Xiao2006A, Cattivelli2008DiffusionRLS, Ber2011Diffusion, Ara2014Adaptive, Vah2017AnalysisRLS, Ras2019Reduced, Yu2019Robust}, and distributed KF \cite{Carli2008Distributed, Saber2009Kalman, Bat2014Kullback, Bat2015Consensus, Li2015Distributed, Das2017Consensus, Ma2017Gossip, Liu2018On, Bat2018Distributed, Zhou2019Distributed, Li2019Distributed, He2020Distributed, Vah2020Performance, Yang2020Distributed, Xie2020Performance}. In our recent work (see e.g. \cite{Xie2018Auto, Xie2018SIAM, Xie2018TAC}), we have established the stability and performance results for the distributed LMS and LS filters, without imposing the usual independence and stationarity assumptions for the system signals. Since the KF algorithm would be optimal when the noise and the parameter variation are white Gaussian noises, here we focus on the KF algorithm in this work. Another reason for us to study this problem is that the existing convergence theory in the literature is far from satisfactory since it can hardly be applied to non-independent and non-stationary signals coming from practical complex systems where feedback loops inevitably exist, and much effort has been devoted to the investigation of distributed KF where the observation matrices of the system are deterministic. 

For examples, \cite{Carli2008Distributed} studied a distributed KF based on consensus strategies, and  \cite{Saber2009Kalman} introduced a scalable suboptimal Kalman-Consensus filter and provided a formal stability and performance analysis. Moreover, \cite{Bat2014Kullback} proposed a distributed a distributed KF algorithm based on covariance intersection method, and analyzed the stability properties, and  \cite{Bat2015Consensus} designed the optimal consensus and innovation gain matrices yielding distributed estimates with minimized mean-squared error.  A quantized gossip-based interactive Kalman Filtering (QGIKF) algorithm for deterministic fixed observation matrices was studied in \cite{Li2015Distributed}, together with the weak convergence. In addition, \cite{Das2017Consensus} developed a Kalman filter type consensus + innovations distributed linear estimator, and designed the optimal consensus and innovation gain matrices yielding distributed estimates with minimized mean-squared error, and \cite{Ma2017Gossip} proposed a gossip-based distributed Kalman filter (GDKF) for deterministic time-varying observation matrices, and provided the error reduction rate. Furthermore, \cite{Liu2018On} and \cite{Bat2018Distributed} considered Kalman-consensus filter for linear time-invariant systems, where the communication links are subject to random failures. A distributed Kalman filtering algorithm of a linear time-invariant discrete-time system in the presence of data packet drops was studied in \cite{Zhou2019Distributed}, and a distributed Kalman filtering for deterministic time-varying observation matrices with mild assumption on communication topology and local observability was studied in \cite{Li2019Distributed}. Moreover, \cite{Vah2020Performance} studied  the performance of partial diffusion Kalman filtering (PDKF) algorithm for the networks with noisy links, and \cite{Yang2020Distributed} designed a distributed Kalman filtering algorithm, where the communication links of the sensor networks are subject to bounded time-varying transmission delays. Furthermore, \cite{He2020Distributed} established the boundedness of the error covariance matrix and the exponentially asymptotic unbiasedness of the state estimate for deterministic time-varying observation matrices. 

To the best of our knowledge, the first step to consider distributed KF algorithms for the dynamical system with general random coefficients is made in \cite{Xie2020Performance}, where each estimator shares local innovation pairs with its neighbors to collectively finish the estimation task. However, the proposed distributed KF algorithm requires to exchange a lot of information since it needs to diffuse $L$ times for each time iteration, where $L$ is not smaller than the diameter of the network topology which increases as the network grows.

In this paper, we will consider a well-known distributed time-varying stochastic linear regression model, and provide a theoretical analysis for a distributed KF algorithm of diffusion type \cite{Bat2014Kullback, Bat2015Consensus, Liu2018On, He2020Distributed} where the diffusion strategy is designed via the so called covariance intersection fusion rule. Each node is only allowed to communicate with its neighbors, and both the estimates of the unknown parameter and the inverse of the covariance matrices are diffused between neighboring nodes in such a diffusion strategy. Note also that it only needs to diffuse one time for each time iteration, which greatly reduces the communication complexity compared with \cite{Xie2020Performance}. The main contributions of the paper contain the following aspects: 1) The stability of the proposed distributed KF algorithm can be obtained without relying on the assumptions of the independency and stationarity of the regression signals, which makes it possible for applications to the stochastic feedback system. 2) The stability result of the proposed distributed KF algorithm is established under a  cooperative excitation condition, which is a natural extension of the single sensor case, and implies that the whole sensor network can accomplish the estimation task cooperatively, even if none of the sensors can do it individually due to lack of sufficient information. 

In the rest of the paper, we will present the graph theory, observation model, and the distributed KF algorithm in Section II. The error equations, mathematical definitions, and assumptions are stated in Section III. The main results and proofs are given in Sections IV and V, respectively. Section VI gives a simulation result and Section VII concludes the paper and discusses related future problems.

\textbf{Basic notations:} In the sequel, a vector $X\in\mathbb{R}^{n}$ is viewed as an $n$-dimensional column real vector and $A\in\mathbb{R}^{m\times n}$ is viewed as an $m\times n$-dimensional real matrix throughout the paper. Let $A\in\mathbb{R}^{n\times n}$ and $B\in\mathbb{R}^{n\times n}$ be two symmetric matrices, then $A\geq B$ means $A-B$ is a positive semidefinite matrix, and $A>B$ means $A-B$ is a positive definite matrix. Let also $\lambda_{max}\{\cdot\}$ and $\lambda_{min}\{\cdot\}$ denote the largest and the smallest eigenvalues of the matrix, respectively. For any matrix $X\in\mathbb{R}^{m\times n}$, the Euclidean norm is defined as $\parallel X\parallel=(\lambda_{max}\{XX^{\top}\})^{\frac{1}{2}}$, where $(\cdot)^{\top}$ denotes the transpose operator. We use $\mathbb{E}[\cdot]$ to denote the mathematical expectation operator, and $\mathbb{E}[\cdot|\mathcal{F}_{k}]$ to denote the conditional mathematical expectation operator, where $\{\mathcal{F}_{k}\}$ is a sequence of nondecreasing $\sigma$-algebras\cite{Chow1978Probability}. Here we use $\log (\cdot)$ to denote the logarithmic operator based on natural number $e$, $\text{Tr}(\cdot)$ and $\vert\cdot\vert$ to denote the trace and determinant of the matrix, respectively. Note that $\vert\cdot\vert$ should not be confused with the absolute value of a scalar from the context.

\section{Problem Formulation}

\subsection{Graph Theory}

As usual, let us consider a set of $n$ vertexes and model it as a directed graph
$\mathcal{G}=(\mathcal{V},\mathcal{E})$, where $\mathcal{V}=\{1, 2,......, n\}$ is the set of vertexes and $\mathcal{E}\subseteq \mathcal{V}\times \mathcal{V}$ is the set of directed arrows. An arrow $(i,j)$ is considered to be directed from $i$ to $j$, where $j$ is called the head and $i$ is called the tail of the arrow. For a vertex, the number of head ends adjacent to a vertex is called the indegree of the vertex, and the number of tail ends adjacent to a vertex is its outdegree. A path of length $\ell$ in the graph $\mathcal{G}$ is a sequence of nodes $\{i_{1},\dots,i_{\ell}\}$ subject to $(i_{j},i_{j+1})\in\mathcal{E}$, for $1\leq j\leq\ell-1$. The distance from vertexes $i$ to $j$ is the minimum value of the length of all the paths from $i$ to $j$, and the diameter of the graph $\mathcal{G}$ is the maximum value of the distances between any two nodes in the graph $\mathcal{G}$.

The structure of the directed graph $\mathcal{G}$ is usually described by a weighted adjacency matrix $\mathcal{A}=\{a_{ij}\}_{n\times n}$, where $a_{ij}>0$ if the arrow $(i, j)\in\mathcal{E}$, which means that $j$ is the head and $i$ is the tail, and $a_{ij}=0$ otherwise. Note that $\mathcal{G}$ is called a balanced digraph, if  $\sum_{j=1}^{n}a_{ji}=\sum_{j=1}^{n}a_{ij}=1,\forall i=1,\dots,n$. In this paper, we assume that the graph $\mathcal{G}$ is balanced. Note also that the matrix $\mathcal{A}$ is asymmetric.

We use vertex $i$ to denote the $i$th sensor and edge $(i,j)$ to denote the communication from sensor $i$ to sensor $j$. Note that $(i,j)\in\mathcal{E}\Leftrightarrow a_{ij}>0$. The set of neighbors of sensor $i$ is denoted as
$$
\mathcal{N}_{i}=\{\ell\in V | (\ell,i)\in \mathcal{E}\},
$$
and any neighboring sensors have the ability to transmit information over the directed arrow between them. 

\subsection{Observation Model}

Let us consider the following time-varying stochastic linear regression model at sensor $i (i=1,\dots,n)$
\begin{equation}\label{model}
y_{k,i}=\bm{\varphi}_{k,i}^{\top}\bm{\theta}_{k}+v_{k,i}, ~~~~k\geq 0,
\end{equation}
where $y_{k,i}\in\mathbb{R}$ and $v_{k,i}\in\mathbb{R}$ are the observation and noise of sensor $i$ at time $k$ respectively, $\bm{\varphi}_{k,i}\in\mathbb{R}^{m}$ is the  stochastic regressor of sensor $i$ at time $k$, and $\bm{\theta}_{k}\in\mathbb{R}^{m}$ is the unknown time-varying parameter which needs to be estimated by each sensor $i$ in the network. Note that the observation matrix $\bm{\varphi}_{k,i}$ in (\ref{model}) is stochastic, while most literature \cite{Carli2008Distributed, Saber2009Kalman, Bat2014Kullback, Bat2015Consensus, Li2015Distributed, Das2017Consensus, Ma2017Gossip, Liu2018On, Bat2018Distributed, Zhou2019Distributed, Li2019Distributed, He2020Distributed, Vah2020Performance, Yang2020Distributed} considered deterministic observation matrix.

In order to develop a strategy to update the estimation of the $m\times 1$-dimensional system signals or time-varying parameter vector $\bm{\theta}_{k}$ in real-time, it is usually convenient to denote the variation of $\bm{\theta}_{k}$ as follows
\begin{equation}\label{iteration}
\bm{\delta}_{k}=\bm{\theta}_{k}-\bm{\theta}_{k-1}, ~~~~k\geq 1,
\end{equation}
where $\bm{\delta}_{k}$ is an undefined $m\times 1$-dimensional vector. Note that (\ref{iteration}) is an simplified system model compare with the linear time-invariant system model considered in \cite{Carli2008Distributed, Saber2009Kalman, Bat2014Kullback, Bat2015Consensus, Li2015Distributed, Das2017Consensus, Ma2017Gossip, Liu2018On, Bat2018Distributed, Zhou2019Distributed, Li2019Distributed, He2020Distributed, Vah2020Performance, Yang2020Distributed, Xie2020Performance}. This is the first step for us to consider distributed KF algorithms for the dynamical system with  random coefficients, i.e., $\bm{\varphi}_{k,i}$. The general linear time-invariant system model will be considered in a future work.

Tracking or estimating a time-varying signal is a fundamental problem in system identification and signal processing, and a variety of recursive algorithms have been derived in the literature \cite{Widerow1985Adaptive, Guo1990Estimating, Guo1994Stability, Solo1995Adaptive, Macchi1995Adaptive, Haykin1996Adaptive}, which usually have the following form:
$$
\widehat{\bm{\theta}}_{k+1,i}=\widehat{\bm{\theta}}_{k,i}+\bm{L}_{k,i}(y_{k,i}-\bm{\varphi}_{k,i}^{\top}\widehat{\bm{\theta}}_{k,i}),
$$
where $\bm{L}_{k,i}$ is the adaptation gain which does not tends to zero as time instant $k$ tends to infinity. This is because when the unknown parameter is time-varying, the algorithm should be persistently alert to follow the parameter variations. The three most common ways of selecting $\bm{L}_{k,i}$ can obtain LMS, RLS and KF algorithms, in which the KF algorithm would be optimal if the noise and the parameter variation are white Gaussian noises. Thus, we focus on the KF algorithm in this work.

Here we first present the traditional non-cooperative KF algorithm as follows. For any given sensor $i=1\dots,n$, begin with an initial estimate $\widehat{\bm{\theta}}_{0,i}\in\mathbb{R}^{m}$, and an initial matrix $P_{0,i}\in\mathbb{R}^{m\times m}$. The KF algorithm is recursively defined for iteration $k\geq 1$ as follows:
\begin{equation}\label{A1}
\widehat{\bm{\theta}}_{k+1,i}=\widehat{\bm{\theta}}_{k,i}+\frac{P_{k,i}\bm{\varphi}_{k,i}}{r_{i}+\bm{\varphi}_{k,i}^{\top}P_{k,i}\bm{\varphi}_{k,i}}(y_{k,i}-\bm{\varphi}_{k,i}^{\top}\widehat{\bm{\theta}}_{k,i}),
\end{equation}
\begin{equation}\label{A2}
P_{k+1,i}=P_{k,i}-\frac{P_{k,i}\bm{\varphi}_{k,i}\bm{\varphi}_{k,i}^{\top}P_{k,i}}{r_{i}+\bm{\varphi}_{k,i}^{\top}P_{k,i}\bm{\varphi}_{k,i}}+Q,
\end{equation}
where $r_{i}\in\mathbb{R}, Q\in\mathbb{R}^{m\times m}$ may be regarded as the priori estimates for the variances of $v_{k,i}$ and $\bm{\delta}_{k}$, and $r_{i}>0, Q>0$ holds. Note that taking $r_{i}$ and $Q$ as constants is just for simplicity of discussion, and generalizations to time-varying cases are straightforward.

To the best of our knowledge, the best result which guarantees the stability of the KF algorithm for each sensor $i$ in the network and allows $\{\bm{\varphi}_{k,i}\}$ to be a large class of stochastic processes appears in the work of Guo \cite{Guo1994Stability}, where it is assumed that $\{\bm{\varphi}_{k,i}, \mathcal{F}_{k,i}\}$ is an adapted process ($\mathcal{F}_{k,i}$ is any family of non-decreasing $\sigma$-algebras) satisfying there exists an integer $h>0$ such that $\{\lambda_{k,i}, k\geq 0\}\in\mathcal{S}^{0}(\lambda)$ for some $\lambda\in(0,1)$, where $\lambda_{k,i}$ is defined by
\begin{equation}\label{KFCON}
\lambda_{k,i}\overset{\triangle}{=}\lambda_{min}\bigg\{\mathbb{E}\bigg[\frac{1}{h+1}\sum_{j=kh+1}^{(k+1)h}\frac{\bm{\varphi}_{j,i}\bm{\varphi}_{j,i}^{\top}}{1+\|\bm{\varphi}_{j,i}\|^{2}}\Big\vert \mathcal{F}_{kh,i}\bigg]\bigg\},
\end{equation}
and $\mathcal{S}^{0}(\lambda)$ is defined in \emph{Definition 3.3}. Note that for high-dimensional or sparse stochastic regressors $\bm{\varphi}_{k,i}$, the condition (\ref{KFCON}) may indeed be not satisfied. This situation may be improved by exchanging information among nodes in a sensor network on which the distributed KF is defined in the following part.

\subsection{Distributed KF Algorithm}

In the following, we present the distributed KF algorithm based on a diffusion strategy. Note that the diffusion strategy is designed via the so called covariance intersection fusion rule as used in e.g., \cite{Bat2014Kullback, Bat2015Consensus, Liu2018On, He2020Distributed}, and the following algorithm can be derived from some existing literature for distributed Kalman filters \cite{Bat2014Kullback, Bat2015Consensus, Liu2018On, He2020Distributed} by assuming that the observation and state equations are (\ref{model}) and  (\ref{iteration}), respectively. The main contribution of this work is to provide the theoretical analysis of the distributed KF algorithm under non-independent and non-stationary signal assumptions.

\begin{algorithm}
\caption{Distributed KF algorithm}
For any given sensor $i=1\dots,n$, begin with an initial estimate $\bm{\theta}_{0,i}\in\mathbb{R}^{m}$, and an initial matrix $P_{0,i}\in\mathbb{R}^{m\times m}$. The algorithm is recursively defined for iteration $k\geq 1$ as follows:

\begin{algorithmic}[1]
\State Adapt:
\begin{equation}\label{A3}
\bar{\bm{\theta}}_{k+1,i}=\widehat{\bm{\theta}}_{k,i}+\frac{P_{k,i}\bm{\varphi}_{k,i}}{r_{i}+\bm{\varphi}_{k,i}^{\top}P_{k,i}\bm{\varphi}_{k,i}}(y_{k,i}-\bm{\varphi}_{k,i}^{\top}\widehat{\bm{\theta}}_{k,i}),
\end{equation}
\begin{equation}\label{A4}
\bar{P}_{k+1,i}=P_{k,i}-\frac{P_{k,i}\bm{\varphi}_{k,i}\bm{\varphi}_{k,i}^{\top}P_{k,i}}{r_{i}+\bm{\varphi}_{k,i}^{\top}P_{k,i}\bm{\varphi}_{k,i}}+Q,
\end{equation}

\State Combine:
\begin{equation}\label{A5}
P_{k+1,i}^{-1}=\sum_{\ell\in\mathcal{N}_{i}}a_{\ell i}\bar{P}_{k+1,\ell}^{-1},
\end{equation}
\begin{equation}\label{A6}
\widehat{\bm{\theta}}_{k+1,i}=P_{k+1,i}\sum_{\ell\in\mathcal{N}_{i}}a_{\ell i}\bar{P}_{k+1,\ell}^{-1}\bar{\bm{\theta}}_{k+1,\ell},
\end{equation}
\end{algorithmic}
where $r_{i}\in\mathbb{R}, Q\in\mathbb{R}^{m\times m}$ may be regarded as the priori estimates for the variances of $v_{k,i}$ and $\bm{\delta}_{k}$, and $r_{i}>0, Q>0$ holds.
\end{algorithm}

Note that when $\mathcal{A}=I_{n}$, the above distributed KF algorithm will degenerate to the non-cooperative KF algorithm. 

\section{Some Preliminaries}

\subsection{Error Equation for the Distributed KF Algorithm}

In order to analyze the above algorithm, we first need to derive the estimation error equation. Then for sensor $i$, define the following two estimation errors:
$$
\begin{aligned}
&\widetilde{\bm{\theta}}_{k,i}=\bm{\theta}_{k}-\widehat{\bm{\theta}}_{k,i},\\
&\widetilde{\bar{\bm{\theta}}}_{k,i}=\bm{\theta}_{k}-\bar{\bm{\theta}}_{k,i}.
\end{aligned}
$$

Then from (\ref{A5}) and (\ref{A6}), we have
\begin{align}\label{error1}
\widetilde{\bm{\theta}}_{k+1,i}=&\bm{\theta}_{k+1}-P_{k+1,i}\sum_{\ell\in\mathcal{N}_{i}}a_{\ell i}\bar{P}_{k+1,\ell}^{-1}\bar{\bm{\theta}}_{k+1,\ell}\nonumber\\
=&P_{k+1,i}\sum_{\ell\in\mathcal{N}_{i}}a_{\ell i}\bar{P}_{k+1,\ell}^{-1}\bm{\theta}_{k+1}\nonumber\\
&-P_{k+1,i}\sum_{\ell\in\mathcal{N}_{i}}a_{\ell i}\bar{P}_{k+1,\ell}^{-1}\bar{\bm{\theta}}_{k+1,\ell}\nonumber\\
=&P_{k+1,i}\sum_{\ell\in\mathcal{N}_{i}}a_{\ell i}\bar{P}_{k+1,\ell}^{-1}\widetilde{\bar{\bm{\theta}}}_{k+1,\ell}.
\end{align}

From (\ref{model}), (\ref{iteration}) and (\ref{A3}), we can also obtain that
\begin{align}\label{error2}
\widetilde{\bar{\bm{\theta}}}_{k+1,i}=&\bm{\theta}_{k+1}-\bar{\bm{\theta}}_{k+1,i}\nonumber\\
=&\bm{\theta}_{k}+\bm{\delta}_{k+1}-\widehat{\bm{\theta}}_{k,i}\nonumber\\
&-\frac{P_{k,i}\bm{\varphi}_{k,i}}{r_{i}+\bm{\varphi}_{k,i}^{\top}P_{k,i}\bm{\varphi}_{k,i}}(y_{k,i}-\bm{\varphi}_{k,i}^{\top}\widehat{\bm{\theta}}_{k,i})\nonumber\\
=&\widetilde{\bm{\theta}}_{k,i}+\bm{\delta}_{k+1}\nonumber
\end{align}
\begin{align}
&-\frac{P_{k,i}\bm{\varphi}_{k,i}}{r_{i}+\bm{\varphi}_{k,i}^{\top}P_{k,i}\bm{\varphi}_{k,i}}(\bm{\varphi}_{k,i}^{\top}\bm{\theta}_{k}-\bm{\varphi}_{k,i}^{\top}\widehat{\bm{\theta}}_{k,i}+v_{k,i})\nonumber\\
=&\Bigg(I_{m}-\frac{P_{k,i}\bm{\varphi}_{k,i}\bm{\varphi}_{k,i}^{\top}}{r_{i}+\bm{\varphi}_{k,i}^{\top}P_{k,i}\bm{\varphi}_{k,i}\bm{\varphi}_{k,i}^{\top}}\Bigg)\widetilde{\bm{\theta}}_{k,i}\nonumber\\
&-\frac{P_{k,i}\bm{\varphi}_{k,i}}{r_{i}+\bm{\varphi}_{k,i}^{\top}P_{k,i}\bm{\varphi}_{k,i}}v_{k,i}+\bm{\delta}_{k+1}.
\end{align}
Denote
$$
L_{k,i}=\frac{P_{k,i}\bm{\varphi}_{k,i}}{r_{i}+\bm{\varphi}_{k,i}^{\top}P_{k,i}\bm{\varphi}_{k,i}},
$$
then we can obtain that
\begin{align}\label{error3}
\widetilde{\bar{\bm{\theta}}}_{k+1,i}=(I_{m}-L_{k,i}\bm{\varphi}_{k,i}^{\top})\widetilde{\bm{\theta}}_{k,i}-L_{k,i}v_{k,i}+\bm{\delta}_{k+1}.
\end{align}
For convenience of analysis, we introduce the following notations:
$$
\begin{aligned}
&\bm{Y}_{k}\overset{\triangle}{=}\text{col}\{y_{k,1},\dots,y_{k,n}\},~~~~~~~~~~~~~~~~~~~~~(n\times 1)\\
&\bm{\Phi}_{k}\overset{\triangle}{=}\text{diag}\{\bm{\varphi}_{k,1},\dots,\bm{\varphi}_{k,n}\},~~~~~~~~~~~~~~~~~~(mn\times n)\\
&\bm{V}_{k}\overset{\triangle}{=}\text{col}\{v_{k,1},\dots,v_{k,n}\},~~~~~~~~~~~~~~~~~~~~~~(n\times 1)\\
&\bm{\Theta}_{k}\overset{\triangle}{=}\text{col}\{\underbrace{\bm{\theta}_{k},\dots,\bm{\theta}_{k}}_{n}\},~~~~~~~~~~~~~~~~~~~~~~~~~(mn\times 1)\\
&\bm{\Delta}_{k}\overset{\triangle}{=}\text{col}\{\underbrace{\bm{\delta}_{k},\dots,\bm{\delta}_{k}}_{n}\},~~~~~~~~~~~~~~~~~~~~~~~~~(mn\times 1)\\
&\widehat{\bm{\Theta}}_{k}\overset{\triangle}{=}\text{col}\{\widehat{\bm{\theta}}_{k,1},\dots,\widehat{\bm{\theta}}_{k,n}\},~~~~~~~~~~~~~~~~~~~~~(mn\times 1)\\
&\bar{\bm{\Theta}}_{k}\overset{\triangle}{=}\text{col}\{\bar{\bm{\theta}}_{k,1},\dots,\bar{\bm{\theta}}_{k,n}\},~~~~~~~~~~~~~~~~~~~~~(mn\times 1)\\
&\widetilde{\bm{\Theta}}_{k}\overset{\triangle}{=}\text{col}\{\widetilde{\bm{\theta}}_{k,1},\dots,\widetilde{\bm{\theta}}_{k,n}\},~~~~~~~~~~~~~~~~~~~~~(mn\times 1)\\
&~~~~~~~\text{where}~ \widetilde{\bm{\theta}}_{k,i}=\bm{\theta}_{k}-\widehat{\bm{\theta}}_{k,i},\\
&\widetilde{\bar{\bm{\Theta}}}_{k}\overset{\triangle}{=}\text{col}\{\widetilde{\bar{\bm{\theta}}}_{k,1},\dots,\widetilde{\bar{\bm{\theta}}}_{k,n}\},~~~~~~~~~~~~~~~~~~~~~(mn\times 1)\\
&~~~~~~~\text{where}~ \widetilde{\bar{\bm{\theta}}}_{k,i}=\bm{\theta}_{k}-\bar{\bm{\theta}}_{k,i},\\
&\bm{L}_{k}\overset{\triangle}{=}\text{diag}\{L_{k,1},\dots,L_{k,n}\},~~~~~~~~~~~~~~~~~~~~(mn\times n)\\
&\bm{P}_{k}\overset{\triangle}{=}\text{diag}\{P_{k,1},\dots,P_{k,n}\},~~~~~~~~~~~~~~~~~~~~(mn\times mn)\\
&\bar{\bm{P}}_{k}\overset{\triangle}{=}\text{diag}\{\bar{P}_{k,1},\dots,\bar{P}_{k,n}\},~~~~~~~~~~~~~~~~~~~~(mn\times mn)\\
&\bm{Q}_{diag}\overset{\triangle}{=}\text{diag}\{\underbrace{Q,\dots,Q}_{n}\},~~~~~~~~~~~~~~~~~~~~~~(mn\times mn)\\
&\bm{\mathscr{A}}\overset{\triangle}{=}\mathcal{A}\otimes I_{m},~~~~~~~~~~~~~~~~~~~~~~~~~~~~~~~~~~~~(mn\times mn)
\end{aligned}
$$
where $\text{col}\{\cdots\}$ denotes a vector by stacking the specified vectors, $\text{diag}\{\cdots\}$ is used in a non-standard manner which means that $m\times 1$ column vectors are combined ``in a diagonal manner'' resulting in a $mn\times n$ matrix, $\mathcal{A}$ is the adjacency matrix, and $\otimes$ is the Kronecker product. Note also that $\bm{\Theta}$ means just the $n$-times replication of vectors $\bm{\theta}$. By (\ref{model}) and (\ref{iteration}), we have
\begin{equation}\label{Model}
\bm{Y}_{k}=\bm{\Phi}_{k}^{\top}\bm{\Theta}_{k}+\bm{V}_{k},
\end{equation}
and
\begin{equation}\label{Iteration}
\bm{\Delta}_{k}=\bm{\Theta}_{k}-\bm{\Theta}_{k-1}, ~~~~k\geq 1,
\end{equation}
For the distributed KF algorithm, we have
\begin{equation}\label{7}
\begin{cases}
&\bar{\bm{\Theta}}_{k+1}=\widehat{\bm{\Theta}}_{k}+\bm{L}_{k}(\bm{Y}_{k}-\bm{\Phi}_{k}^{\top}\widehat{\bm{\Theta}}_{k}),\\
&\bar{\bm{P}}_{k+1}=\bm{P}_{k}-\bm{L}_{k}\bm{\Phi}_{k}^{\top}\bm{P}_{k}+\bm{Q}_{diag},\\
&\text{vec}\{\bm{P}_{k+1}^{-1}\}=\mathscr{A}^{\top}\text{vec}\{\bar{\bm{P}}_{k+1}^{-1}\},\\
&\bm{\Theta}_{k+1}=\bm{P}_{k+1}\mathscr{A}^{\top}\bar{\bm{P}}_{k+1}^{-1}\bar{\bm{\Theta}}_{k+1},
\end{cases}
\end{equation}
where $\text{vec}\{\cdot\}$ denotes the operator that stacks the blocks of a block diagonal matrix on top of each other. Since $\widetilde{\bm{\Theta}}_{k}=\bm{\Theta}-\bm{\Theta}_{k}$ and $\widetilde{\bar{\bm{\Theta}}}_{k}=\bm{\Theta}-\bar{\bm{\Theta}}_{k}$, we can get from (\ref{error3}) that
$$
\widetilde{\bar{\bm{\Theta}}}_{k+1}=(I_{mn}-\bm{L}_{k}\bm{\Phi}_{k}^{\top})\widetilde{\bm{\Theta}}_{k}-\bm{L}_{k}\bm{V}_{k}+\bm{\Delta}_{k+1},
$$
and by (\ref{error1}), we have
\begin{align}
\widetilde{\bm{\Theta}}_{k+1}=&\bm{P}_{k+1}\mathscr{A}^{\top}\bar{\bm{P}}_{k+1}^{-1}\widetilde{\bar{\bm{\Theta}}}_{k+1}\nonumber\\
=&\bm{P}_{k+1}\mathscr{A}^{\top}\bar{\bm{P}}_{k+1}^{-1}(I_{mn}-\bm{L}_{k}\bm{\Phi}_{k}^{\top})\widetilde{\bm{\Theta}}_{k}\nonumber\\
&-\bm{P}_{k+1}\mathscr{A}^{\top}\bar{\bm{P}}_{k+1}^{-1}\bm{L}_{k}\bm{V}_{k}\nonumber\\
&+\bm{P}_{k+1}\mathscr{A}^{\top}\bar{\bm{P}}_{k+1}^{-1}\bm{\Delta}_{k+1}.
\end{align}

In the following section, we will analyze the stability of the above distributed KF algorithm under non-independent and correlated signal assumptions.

\subsection{Some definitions}

We use $\mathcal{F}_{k}=\sigma\{\bm{\varphi}_{i}^{j},\bm{\omega}_{i},v_{i-1}^{j},j=1,\dots,n,i\leq k\}$ to denote the $\sigma$-algebra generated by $\{\bm{\varphi}_{i}^{j}, \bm{\omega}_{i}, v_{i-1}^{j}, j=1,\dots,n,i\leq k\}$. To proceed with further discussions, we need the following definitions introduced in \cite{Guo1994Stability}.

\begin{defn}
For a random matrix sequence $\{A_{k},k\geq 0\}$ defined on the basic probability space $(\Omega,\mathcal{F},P)$,
if
$$
\sup_{k\geq 0}\mathbb{E}[\parallel A_{k}\parallel^{p}]<\infty,
$$
holds for some $p>0$, then $\{A_{k}\}$ is called $L_{p}$-bounded. Furthermore, if $\{A_{k}\}$ is a solution of a random difference equation, then $\{A_{k}\}$ is called $L_{p}$-stable.
\end{defn}

\begin{defn}
For a sequence of $s\times s$ random matrices $A=\{A_{k},k\geq 0\}$, if it belongs to the following set with $p\geq 0$,
\begin{align}\label{Sp}
S_{p}(\lambda)=\bigg\{A:&\bigg\|\prod_{j=i+1}^{k}(I-A_{j})\bigg\|_{L_{p}}\leq M\lambda^{k-i},\nonumber\\
&\forall k\geq i+1,\forall i\geq 0, \text{for some} ~M>0\bigg\},
\end{align}
then $\{I-A_{k},k\geq 0\}$ is called $L_{p}$-exponentially stable with parameter $\lambda\in [0,1)$.
\end{defn}

\begin{rem}
As pointed out in literature \cite{Guo1994Stability}, (\ref{Sp}) is in some sense the necessary and sufficient condition for stability of random linear equations of the form $x_{k} = (I-A_{k})x_{k} + \xi_{k+1}, k\geq 0$, and it is well known that the analysis of such a random matrix product is a mathematically difficult problem. However, as demonstrated by Guo \cite{Guo1994Stability}, for linear random equations arising from adaptive filtering algorithms,  it is possible to transfer the product of the random matrices to that of a certain class of scalar sequences, and the later can be further analyzed based on some excitation or information conditions on the regressors. To this end, we introduce the following subclass of $S_{1}(\lambda)$.
\end{rem}

\begin{defn}
For a scalar sequence $a=\{a_{k},k\geq 0\}$ and $\lambda\in(0,1)$, we set
\begin{align}\label{S0}
S^{0}(\lambda)=\bigg\{a:&a_{k}\in[0,1],\mathbb{E}\bigg[\prod_{j=i+1}^{k}(1-a_{j})\bigg]\leq M\lambda^{k-i},\nonumber\\
&\forall k\geq i+1,\forall i\geq 0, \text{for some} ~M>0\bigg\}.
\end{align}
\end{defn}

\begin{rem}
This definition will be used to introduce the cooperative information condition in the following part.
\end{rem}

\subsection{Assumptions}

In order to guaranteer the stability of the distributed KF algorithm, the following network topology assumption is naturally required to avoid isolated nodes in the network.

\begin{asm}\label{asmtopology}
\textbf{(Network Topology)} The digraph $\mathcal{G}$ is strongly connected\footnote{There exists a path between any two vertices in the digraph.} and balanced.
\end{asm}

\begin{rem}
By \emph{Assumption \ref{asmtopology}} and \cite{Horn2013Matrix}, we know that when $\ell$ is no less than the diameter of the graph $\mathcal{G}$, i.e., $D_{\mathcal{G}}$, each entry of the matrix $\mathcal{A}^{\ell}$ shall be positive.
\end{rem}

\begin{asm}\label{asminformation}
\textbf{(Cooperative Information Condition)} For the  adapted sequences $\{\bm{\varphi}_{k,i}, \mathcal{F}_{k}, k\geq 0\} (i=1,\dots,n)$, where $\mathcal{F}_{k}=\sigma\{\bm{\varphi}_{j,i},\bm{\delta}_{j},v_{j-1,i}, i=1,\dots,n, j\leq k\}$, there exists an integer $h>0$ such that $\{\lambda_{k},k\geq 0\}\in S^{0}(\lambda)$ for some $\lambda\in(0,1)$, where $\lambda_{k}$ is defined by
\begin{equation}\label{lambdak}
\lambda_{k}\overset{\triangle}{=}\lambda_{min}\bigg\{\mathbb{E}\bigg[\frac{1}{n(h+1)}\sum_{i=1}^{n}\sum_{j=kh+1}^{(k+1)h}\frac{\bm{\varphi}_{j,i}(\bm{\varphi}_{j,i})^{\top}}{1+\|\bm{\varphi}_{j,i}\|^{2}}\Big\vert \mathcal{F}_{kh}\bigg]\bigg\},
\end{equation}
where $\mathbb{E}[\cdot\vert\mathcal{F}_{kh}]$ is the conditional mathematical expectation operator.
\end{asm}

\begin{rem}
Almost all the existing literature on the theoretical analyses of distributed adaptive filters requires some stringent conditions on the regressors, such as independency and stationarity, which cannot be satisfied for signals generated from stochastic feedback systems. In fact, \emph{Assumption \ref{asminformation}} is a natural generalization of the information condition from single sensor to sensor networks, which is not independent or stationary signal conditions. This conditional mathematical expectation-based information condition for single sensor case was first introduced by Guo in \cite{Guo1990Estimating} and then refined in \cite{Guo1994Stability}, which is quite general for exponential stability (see \cite{Guo1994Stability}) of the adaptive filtering algorithms. Note that \emph{Assumption \ref{asminformation}} implies that the regressor signals will have some kind of ``persistent excitations'' since the prediction of the ``future'' is non-degenerate given the ``past'', which is required to track constantly changing unknown signals. Moreover, under \emph{Assumption \ref{asminformation}}, the distributed KF algorithm can be shown to have the capability to fulfil the estimation task cooperatively even if any sensor cannot estimate the unknown signal individually.
\end{rem}

\section{The Main Results}

By (\ref{7}), we have
\begin{align}
\bar{\bm{P}}_{k+1}=&(I_{mn}-\bm{L}_{k}\bm{\Phi}_{k}^{\top})\bm{P}_{k}(I_{mn}-\bm{L}_{k}\bm{\Phi}_{k}^{\top})^{\top}\nonumber\\
&+\bm{R}\bm{L}_{k}\bm{L}_{k}^{\top}+\bm{Q}_{diag},
\end{align}
where
$$
\bm{R}\overset{\triangle}{=}\text{diag}\{r_{1},\dots,r_{n}\}\otimes I_{m}.
$$

Denote
\begin{align}
&\bm{A}_{k}\overset{\triangle}{=}\bm{L}_{k}\bm{\Phi}_{k}^{\top},\nonumber\\
&\bm{B}_{k}\overset{\triangle}{=}I-\bm{A}_{k},\nonumber\\
&\bm{Q}_{k}\overset{\triangle}{=}\bm{R}\bm{L}_{k}\bm{L}_{k}^{\top}+\bm{Q}_{diag}.
\end{align}

For non-symmetrical random matrix $\{\bm{A}_{k}\}$, the following lemma transfer the study of $\{\bm{A}_{k}\}$ to that of a scalar random sequence in $S^{0}(\lambda)$.

\begin{thm}\label{thm1}
Let $\{\bm{A}_{k}\}$ be a sequence of $mn\times mn$ random matrices, and $\{\bm{Q}_{k}\}$ be a sequence of positive definite random matrices. Then for $\{\bm{P}_{k}\}$ and $\{\bar{\bm{P}}_{k}\}$ recursively defined by
\begin{align}
\bar{\bm{P}}_{k+1}=(I_{mn}-\bm{A}_{k})\bm{P}_{k}(I_{mn}-\bm{A}_{k})^{\top}+\bm{Q}_{k},
\end{align}
and
\begin{align}
\text{vec}\{\bm{P}_{k}^{-1}\}=\mathscr{A}^{\top}\text{vec}\{\bar{\bm{P}}_{k}^{-1}\},
\end{align}
we have for all $t>s$,
\begin{align}
&\Bigg\|\prod_{k=s}^{t-1}\bm{P}_{k+1}\mathscr{A}^{\top}\bar{\bm{P}}_{k+1}^{-1}(I_{mn}-\bm{A}_{k})\Bigg\|^{2}\nonumber\\
\leq &\Bigg\{\prod_{k=s}^{t-1}\Bigg(1-\frac{1}{1+\|\bm{Q}_{k}^{-1}\bm{\bar{P}}_{k+1}\|}\Bigg)\Bigg\}\Big\{\|\bm{P}_{t}\|\cdot\|\bm{P}_{s}^{-1}\|\Big\}.
\end{align}
Hence if $\{\bm{P}_{k}\}$ satisfies the following two conditions:
\begin{enumerate}
\item
$$
\Bigg\{\frac{1}{1+\|\bm{Q}_{k}^{-1}\bar{\bm{P}}_{k+1}\|}\Bigg\}\in S^{0}(\lambda), ~~for~~some~~\lambda\in[0,1);
$$
\item
$$
\sup_{t\geq s\geq 0}\|(\|\bm{P}_{t}\|\cdot\|\bm{P}_{s}^{-1}\|)\|_{L_{p}}<\infty, ~~for~~some~~p\geq 1,
$$
\end{enumerate}
then
\begin{equation}
\{I_{mn}-\bm{P}_{k+1}\mathscr{A}^{\top}\bar{\bm{P}}_{k+1}^{-1}(I_{mn}-\bm{A}_{k})\}\in S_{p}(\lambda^{1/2p}).
\end{equation}
\end{thm}

The proof of \emph{Theorem \ref{thm1}} is given Section V. We now proceed to analyze the stability of the distributed KF algorithm. Before applying \emph{Theorem \ref{thm1}}, we need to prove some boundedness properties of $\{\bm{P}_{k}\}$ first.

\begin{lem}\label{lem1}
For $\{\bm{P}_{k}\}$ generated by (\ref{7}), if \emph{Assumptions \ref{asmtopology}} and \emph{\ref{asminformation}} are satisfied, then there exists a constant $\varepsilon^{*}$ such that for any $\varepsilon\in[0,\varepsilon^{*})$,
\begin{equation}
\sup_{k\geq 0}\mathbb{E}[\exp(\varepsilon\|\bm{P}_{k}\|)]<\infty.
\end{equation}
\end{lem}

The proof of \emph{Lemma \ref{lem1}} is given in Section V. The following result is a direct consequence of \emph{Lemma \ref{lem1}}. Then we omit the proof here.

\begin{cor}
For $\{\bm{P}_{k}\}$ generated by (\ref{7}), if \emph{Assumptions \ref{asmtopology}} and \emph{\ref{asminformation}} are satisfied, then for any $p>0$,
\begin{equation}
\sup_{k\geq 0}\mathbb{E}[\|\bm{P}_{k}\|^{p}]<\infty.
\end{equation}
\end{cor}

\begin{lem}\label{lem2}
For $\{\bm{P}_{k}\}$ generated by (\ref{7}), if \emph{Assumptions \ref{asmtopology}} and \emph{\ref{asminformation}} are satisfied, then for any $\mu\in(0,1]$, there exists a constant $\lambda$ such that
\begin{equation}
\Bigg\{\frac{\mu}{1+\|\bm{Q}_{diag}^{-1}\|\cdot\|\bar{\bm{P}}_{k}\|}\Bigg\}\in S^{0}(\lambda).
\end{equation}
\end{lem}

The proof of \emph{Lemma \ref{lem2}} is given in Section V. Then we can obtain the following tracking error bound for the distributed KF algorithm.

By the above results, we can obtain the following upper bound for the estimation error.

\begin{thm}\label{thm2}
Consider the observation model (\ref{model}) and the distributed KF algorithm (\ref{7}). Suppose that \emph{Assumptions \ref{asmtopology}} and \emph{\ref{asminformation}} are satisfied and that for some $p\geq 1$ and $\beta>2$,
\begin{equation}
\sigma_{p}\overset{\triangle}{=}\sup_{k}\|\xi_{k}\log^{\beta}(e+\xi_{k})\|_{L_{p}}<\infty,
\end{equation}
and
\begin{equation}
\|\widetilde{\bm{\Theta}}_{0}\|_{L_{2p}}<\infty,
\end{equation}
where $\xi_{k}=\|\bm{V}_{k}\|+\|\bm{\Delta}_{k+1}\|$. Then the tracking error $\{\widetilde{\bm{\Theta}}_{k}, k\geq 0\}$ is $L_{p}$-stable and
$$
\limsup_{k\to\infty}\|\widetilde{\bm{\Theta}}_{k}\|_{L_{p}}\leq c[\sigma_{p}\log^{1+\beta/2}(e+\sigma_{p}^{-1})],
$$
where $c$ is a finite constant depending on $\{\bm{\Phi}_{k}\}, \bm{R}, \bm{Q}_{diag}$ and $p$ only.
\end{thm}

\begin{IEEEproof}
By (\ref{7}), we have
\begin{align}
\bar{\bm{P}}_{k+1}=(I_{mn}-\bm{L}_{k}\bm{\Phi}_{k}^{\top})\bm{P}_{k}(I_{mn}-\bm{L}_{k}\bm{\Phi}_{k}^{\top})^{\top}+\bm{Q}_{k},
\end{align}
where $\bm{Q}_{k}=\bm{R}\bm{L}_{k}\bm{L}_{k}^{\top}+\bm{Q}_{diag}$. It is easy to see that $\bm{Q}_{k}\geq \bm{Q}_{diag}$ and $\bar{\bm{P}}_{k}\geq\bm{Q}_{diag}$, and by the (\ref{7}), we know that
$$
\|\bm{P}_{k}^{-1}\|\leq\|\bar{\bm{P}}_{k}^{-1}\|\leq\|\bm{Q}_{diag}^{-1}\|.
$$

Hence, by \emph{Theorem \ref{thm1}}, we have for all $t>s$,
\begin{align}
&\Bigg\|\prod_{k=s}^{t-1}\bm{P}_{k+1}\mathscr{A}^{\top}\bar{\bm{P}}_{k+1}^{-1}(I_{mn}-\bm{A}_{k})\Bigg\|\nonumber\\
\leq &\Bigg\{\prod_{k=s}^{t-1}\Bigg(1-\frac{1}{1+\|\bm{Q}_{diag}^{-1}\|\cdot\|\bm{\bar{P}}_{k+1}\|}\Bigg)^{\frac{1}{2}}\Bigg\}\nonumber\\
&\cdot\Big\{\|\bm{P}_{t}\|^{\frac{1}{2}}\cdot\|\bm{Q}_{diag}^{-1}\|^{\frac{1}{2}}\Big\}.
\end{align}

Note also that
$$
\|\bm{L}_{k}\|\leq\frac{\|\bm{P}_{k}\|^{\frac{1}{2}}}{2\sqrt{r_{min}}},
$$
and
$$
\|\bm{P}_{k+1}\mathscr{A}^{\top}\bar{\bm{P}}_{k+1}^{-1}\|\leq\|\bm{P}_{k+1}\|\cdot\|\bar{\bm{P}}_{k+1}^{-1}\|\leq\|\bm{P}_{k+1}\|\cdot\|\bm{Q}_{diag}^{-1}\|,
$$
hold, where $r_{min}=\min_{i=1,\dots,n}\{r_{1},\dots,r_{n}\}$. Hence we have
\begin{align}
&\|\widetilde{\bm{\Theta}}_{k+1}\|_{L_{p}}\nonumber\\
\leq&\Bigg\|\prod_{i=0}^{k}\bm{P}_{i+1}\mathscr{A}^{\top}\bar{\bm{P}}_{i+1}^{-1}(I_{mn}-\bm{A}_{i})\widetilde{\bm{\Theta}}_{0}\Bigg\|_{L_{p}}\nonumber\\
&+\|\bm{Q}_{diag}^{-1}\|^{\frac{3}{2}}\sum_{i=0}^{k}\Bigg\|\prod_{j=i+1}^{k}\Bigg(1-\frac{1}{2(1+\|\bm{Q}_{diag}^{-1}\|\cdot\|\bm{\bar{P}}_{j+1}\|)}\Bigg)\nonumber\\
&\cdot\|\bm{P}_{k+1}\|^{\frac{1}{2}}\Bigg(\|\bm{P}_{i}\|+\frac{\|\bm{P}_{i}\|^{\frac{3}{2}}}{2\sqrt{r_{min}}}\Bigg)\xi_{i}\Bigg\|_{L_{p}}\nonumber\\
\leq&\Bigg\|\prod_{i=0}^{k}\bm{P}_{i+1}\mathscr{A}^{\top}\bar{\bm{P}}_{i+1}^{-1}(I_{mn}-\bm{A}_{i})\widetilde{\bm{\Theta}}_{0}\Bigg\|_{L_{p}}\nonumber\\
&+\|\bm{Q}_{diag}^{-1}\|^{\frac{3}{2}}\sup_{i\geq 0}\|\bm{P}_{i}\|_{L_{p}}\nonumber\\
&\cdot\sum_{i=0}^{k}\Bigg\|\prod_{j=i+1}^{k}\Bigg(1-\frac{1}{2(1+\|\bm{Q}_{diag}^{-1}\|\cdot\|\bm{\bar{P}}_{j+1}\|)}\Bigg)\nonumber\\
&\cdot\|\bm{P}_{k+1}\|^{\frac{1}{2}}\Bigg(1+\frac{\|\bm{P}_{i}\|^{\frac{1}{2}}}{2\sqrt{r_{min}}}\Bigg)\xi_{i}\Bigg\|_{L_{p}}
\end{align}

Note that by the Schwarz inequality and \emph{Lemma \ref{lem1}}, we know that
$$
\begin{aligned}
&\sup_{k\geq i}\mathbb{E}[\exp(\varepsilon\|\bm{P}_{k+1}\|^{\frac{1}{2}}\cdot\|\bm{P}_{i}\|^{\frac{1}{2}})]\\
\leq & \sup_{k\geq i}\{\mathbb{E}[\exp(\varepsilon\|\bm{P}_{k+1}\|)]\}^{\frac{1}{2}}\cdot\{\mathbb{E}[\exp(\varepsilon\|\bm{P}_{i}\|)]\}^{\frac{1}{2}}<\infty.
\end{aligned}
$$

By \emph{Lemmas \ref{lem1}} and {\ref{lem2}}, we can see that the following proof can be derived in a similar way as that of Theorem 4.1 in \cite{Guo1994Stability}, details will be omitted here.
\end{IEEEproof}

\begin{rem}
Intuitively, by \emph{Theorem \ref{thm2}} we know that when both the noise and the parameter variation are small, the processes $\xi_k$ and $\sigma_p$ will also be small, and hence the parameter tracking error $\widetilde{\bm{\Theta}}_{k}$ will be small too. Here we only require that the observation noise and the parameter variation satisfy a moment assumption, and no independent, stationary or Gaussian property is required. 
\end{rem}

\section{Proofs of the main results}

\subsection{Proof of \emph{Theorem \ref{thm1}}}

To accomplish the proof of \emph{Theorem \ref{thm1}}, we also need to prove the following lemmas firstly.

\begin{lem}\label{lem51}
For any adjacency matrix $\mathcal{A}=\{a_{ij}\}\in\mathbb{R}^{n\times n}$, denote $\mathscr{A}=\mathcal{A}\otimes I_{m}$, and for any positive definite matrices $Q_{i}\in\mathbb{R}^{m\times m}, i=1,\dots,n$, denote $Q=\text{diag}\{Q_{1},\dots,Q_{n}\}$ and $Q^{'}=\text{diag}\{Q_{1}^{'},\dots,Q_{n}^{'}\}$, where $Q_{i}^{'}=\sum_{j=1}^{n}a_{ji}Q_{j}$. Then the following inequality holds:
\begin{align}\label{AQA}
\mathscr{A}^{\top}Q\mathscr{A}\leq Q^{'}.
\end{align}
\end{lem}

\begin{IEEEproof}
By the definition of $\mathscr{A}$ and $Q$, we can get that
$$
\mathscr{A}^{\top}Q\mathscr{A}=\begin{pmatrix}
\sum\limits_{j=1}^{n}a_{j1}^{2}Q_{j} & \cdots & \sum\limits_{j=1}^{n}a_{j1}a_{jn}Q_{j}\\
\sum\limits_{j=1}^{n}a_{j2}a_{j1}Q_{j} & \cdots & \sum\limits_{j=1}^{n}a_{j2}a_{jn}Q_{j}\\
\vdots & \ddots & \vdots\\
\sum\limits_{j=1}^{n}a_{jn}a_{j1}Q_{j} & \cdots & \sum\limits_{j=1}^{n}a_{jn}^{2}Q_{j}
\end{pmatrix}.
$$

In order to prove (\ref{AQA}), we only need to prove that for any  unit column vector $x\in\mathbb{R}^{mn}$ with $\|x\|=1$, $x^{\top}\mathscr{A}^{\top}Q\mathscr{A}x\leq x^{\top}Q^{'}x$ holds. Here we denote $x=\text{col}\{x_{1}, x_{2}, \dots, x_{n}\}$, where $x_{i}\in\mathbb{R}^{m}$, then by the H\"{o}lder inequality and noticing that $Q_{j}\geq 0 (j=1,\dots,n)$, we have
$$
\begin{aligned}
&x^{\top}\mathscr{A}^{\top}Q\mathscr{A}x\\
=&\sum_{p=1}^{n}\sum_{q=1}^{n}\sum_{j=1}^{n}a_{jp}a_{jq}x_{p}^{\top}Q_{j}x_{q}\\
=&\sum_{p=1}^{n}\sum_{q=1}^{n}\sum_{j=1}^{n}\sqrt{a_{jp}a_{jq}}x_{p}^{\top}Q_{j}^{\frac{1}{2}}\cdot \sqrt{a_{jp}a_{jq}}Q_{j}^{\frac{1}{2}}x_{q}\\
\leq &\Bigg\{\sum_{p=1}^{n}\sum_{q=1}^{n}\sum_{j=1}^{n}a_{jp}a_{jq}x_{p}^{\top}Q_{j}x_{p}\Bigg\}^{\frac{1}{2}}\\
&\cdot\Bigg\{\sum_{p=1}^{n}\sum_{q=1}^{n}\sum_{j=1}^{n}a_{jp}a_{jq}x_{q}^{\top}Q_{j}x_{q}\Bigg\}^{\frac{1}{2}}\\
= &\Bigg\{\sum_{p=1}^{n}\sum_{j=1}^{n}a_{jp}x_{p}^{\top}Q_{j}x_{p}\Bigg\}^{\frac{1}{2}}\Bigg\{\sum_{q=1}^{n}\sum_{j=1}^{n}a_{jq}x_{q}^{\top}Q_{j}x_{q}\Bigg\}^{\frac{1}{2}}\\
=&\sum_{i=1}^{n}\sum_{j=1}^{n}a_{ji}x_{i}^{\top}Q_{j}x_{i}\\
=&x^{\top}Q^{'}x,
\end{aligned}
$$
which completes the proof.
\end{IEEEproof}

\begin{rem}
Note that when $\mathcal{A}=I_{n}$, $Q^{'}-\mathscr{A}^{\top}Q\mathscr{A}=0$ holds. Otherwise, $Q^{'}-\mathscr{A}^{\top}Q\mathscr{A}\geq 0$. By \emph{Lemma \ref{lem51}}, we can obtain the following result.
\end{rem}

\begin{lem}\label{lem52}
For any adjacency matrix $\mathcal{A}=\{a_{ij}\}\in\mathbb{R}^{n\times n}$, denote $\mathscr{A}=\mathcal{A}\otimes I_{m}$. Then for any $k\geq 1$,
\begin{equation}
\mathscr{A}^{\top}\bar{\bm{P}}_{k+1}^{-1}\mathscr{A}\leq \bm{P}_{k+1}^{-1},
\end{equation}
and
\begin{equation}\label{APA}
\mathscr{A}\bm{P}_{k+1}\mathscr{A}^{\top}\leq \bar{\bm{P}}_{k+1},
\end{equation}
holds, where $\bar{\bm{P}}_{k+1}$ and $\bm{P}_{k+1}$ are defined in (\ref{7}).
\end{lem}

\begin{IEEEproof}
By \emph{Lemma \ref{lem51}}, taking $Q_{i}=\bar{P}_{k+1,i}^{-1}\geq 0$, and noting that $P_{k+1,i}^{-1}=\sum_{j=1}^{n}a_{ji}\bar{P}_{k+1,j}^{-1}=Q_{i}^{'}$, we know that
$$
\mathscr{A}^{\top}\bar{\bm{P}}_{k+1}^{-1}\mathscr{A}\leq \bm{P}_{k+1}^{-1},
$$
holds. As for (\ref{APA}), we first assume that $\mathscr{A}$ is invertible, then $0<\mathscr{A}^{\top}\bar{\bm{P}}_{k+1}^{-1}\mathscr{A}\leq \bm{P}_{k+1}^{-1}$ holds. Then by \emph{Lemma \ref{lemA1}} in the Appendix, it is easy to see that
$$
\mathscr{A}\bm{P}_{k+1}\mathscr{A}\leq \bar{\bm{P}}_{k+1}.
$$

Next, we consider the case where $\mathscr{A}$ is not invertible. Since the number of eigenvalues of the matrix $\mathscr{A}$ is finite,  then exists a constant $\varepsilon^{*}\in(0,1)$ such that the perturbed adjacency matrix $\mathscr{A}^{\varepsilon}=\mathscr{A}+\varepsilon I_{mn}=\{a_{ij}^{\varepsilon}\}$ will be invertible for any $0<\varepsilon<\varepsilon^{*}$. By the definition of $\mathscr{A}^{\varepsilon}$, we know that $\mathscr{A}^{\varepsilon}$ is symmetric and the sums of each columns and rows of the matrix $\mathscr{A}^{\varepsilon}$ are all $1+\varepsilon$. Then we define
$$
(P_{k+1,i}^{\varepsilon})^{-1}=\sum_{j=1}^{n}a_{ji}^{\varepsilon}\bar{P}_{k+1,j}^{-1},
$$
and we can denote $\bm{P}_{k+1}^{\varepsilon}=\text{diag}\{P_{k+1,1}^{\varepsilon}, \dots, P_{k+1,n}^{\varepsilon}\}$ since $(P_{k+1,i}^{\varepsilon})^{-1}$ defined above is invertible. Similar to the proof of \emph{Lemma \ref{lem51}}, for any unit column vector $x\in\mathbb{R}^{mn}$, we have
$$
\begin{aligned}
&x^{\top}(\mathscr{A}^{\varepsilon})^{\top}\bar{\bm{P}}_{k+1}^{-1}\mathscr{A}^{\varepsilon}x\\
\leq &\Bigg\{\sum_{p=1}^{n}\sum_{q=1}^{n}\sum_{j=1}^{n}a_{jp}^{\varepsilon}a_{jq}^{\varepsilon}x_{p}^{\top}\bar{P}_{k+1,j}^{-1}x_{p}\Bigg\}^{\frac{1}{2}}\\
&\cdot\Bigg\{\sum_{p=1}^{n}\sum_{q=1}^{n}\sum_{j=1}^{n}a_{jp}^{\varepsilon}a_{jq}^{\varepsilon}x_{q}^{\top}\bar{P}_{k+1,j}^{-1}x_{q}\Bigg\}^{\frac{1}{2}}\\
=&(1+\varepsilon)\sum_{i=1}^{n}\sum_{j=1}^{n}a_{ji}^{\varepsilon}x_{i}^{\top}\bar{P}_{k+1,j}^{-1}x_{i}\\
=&(1+\varepsilon)x^{\top}(\bm{P}_{k+1}^{\varepsilon})^{-1}x.
\end{aligned}
$$
Consequently, we have $(\mathscr{A}^{\varepsilon})^{\top}\bar{\bm{P}}_{k+1}^{-1}\mathscr{A}^{\varepsilon}\leq (1+\varepsilon)(\bm{P}_{k+1}^{\varepsilon})^{-1}$. Since $\mathscr{A}^{\varepsilon}$ is invertible, we know from \emph{Lemma \ref{lem1}} that
$$
\mathscr{A}^{\varepsilon}\bm{P}_{k+1}^{\varepsilon}(\mathscr{A}^{\varepsilon})^{\top}\leq (1+\varepsilon)\bar{\bm{P}}_{k+1}.
$$
By taking $\varepsilon\to 0$ on both sides of the above equation, we can obtain that
$$
\begin{aligned}
&\lim_{\varepsilon\to 0}\mathscr{A}^{\varepsilon}\bm{P}_{k+1}^{\varepsilon}(\mathscr{A}^{\varepsilon})^{\top}=\mathscr{A}\bm{P}_{k+1}\mathscr{A}^{\top}\\
\leq &\lim_{\varepsilon\to 0}(1+\varepsilon)\bar{\bm{P}}_{k+1}=\bar{\bm{P}}_{k+1}.
\end{aligned}
$$
This completes the proof.
\end{IEEEproof}

The proof of \emph{Theorem \ref{thm1}} is given in the following part:

\begin{IEEEproof}
Consider the following equation for $t>s$,
\begin{equation}
\bm{x}_{k+1}=\bm{P}_{k+1}\mathscr{A}^{\top}\bar{\bm{P}}_{k+1}^{-1}(I_{mn}-\bm{A}_{k})\bm{x}_{k},~~~~k\in[s,t-1],
\end{equation}
where $\bm{x}_{s}\in\mathbb{R}^{mn}$ ia taken to be deterministic and $\|\bm{x}_{s}\|=1$. Then
\begin{equation}
\bm{x}_{t}=\prod_{k=s}^{t-1}\bm{P}_{k+1}\mathscr{A}^{\top}\bar{\bm{P}}_{k+1}^{-1}(I_{mn}-\bm{A}_{k})\bm{x}_{s}.
\end{equation}

Next we consider the following Lyapunov function:
$$
V_{k}=\bm{x}_{k}^{\top}\bm{P}_{k}^{-1}\bm{x}_{k}.
$$ 
Denote $\bm{B}_{k}=I-\bm{A}_{k}$, then by \emph{Lemma \ref{lemA1}} in the Appendix and \emph{Lemma \ref{lem52}}, we have
\begin{align}
V_{k+1}=&\bm{x}_{k+1}^{\top}\bm{P}_{k+1}^{-1}\bm{x}_{k+1}\nonumber\\
=&\bm{x}_{k}^{\top}\bm{B}_{k}^{\top}\bar{\bm{P}}_{k+1}^{-1}\mathscr{A}\bm{P}_{k+1}\mathscr{A}^{\top}\bar{\bm{P}}_{k+1}^{-1}\bm{B}_{k}\bm{x}_{k},
\end{align}
and
\begin{align}
&\bm{B}_{k}^{\top}\bar{\bm{P}}_{k+1}^{-1}\mathscr{A}\bm{P}_{k+1}\mathscr{A}^{\top}\bar{\bm{P}}_{k+1}^{-1}\bm{B}_{k}\nonumber\\
\leq &\bm{B}_{k}^{\top}\bar{\bm{P}}_{k+1}^{-1}\bm{B}_{k}\nonumber\\
=&\bm{B}_{k}^{\top}(\bm{B}_{k}\bm{P}_{k}\bm{B}_{k}^{\top}+\bm{Q}_{k})^{-1}\bm{B}_{k}\nonumber\\
=&\bm{P}_{k}^{-1}-(\bm{P}_{k}+\bm{P}_{k}\bm{B}_{k}^{\top}\bm{Q}_{k}^{-1}\bm{B}_{k}\bm{P}_{k})^{-1}\nonumber\\
=&\bm{P}_{k}^{-1/2}(I_{mn}-[I_{mn}+\bm{P}_{k}^{1/2}\bm{B}_{k}^{\top}\bm{Q}_{k}^{-1}\bm{B}_{k}\bm{P}_{k}^{1/2}])\bm{P}_{k}^{-1/2}\nonumber\\
\leq & (1-[1+\|\bm{Q}_{k}^{-1}\bm{B}_{k}\bm{P}_{k}\bm{B}_{k}^{\top}\|]^{-1})\bm{P}_{k}^{-1},
\end{align}
which yields
$$
V_{k+1}\leq\Bigg(1-\frac{1}{1+\|\bm{Q}_{k}^{-1}\bar{\bm{P}}_{k+1}\|}\Bigg)V_{k},
$$
and so
$$
V_{t}\leq\prod_{k=s}^{t-1}\Bigg(1-\frac{1}{1+\|\bm{Q}_{k}^{-1}\bar{\bm{P}}_{k+1}\|}\Bigg)V_{s}.
$$

Hence we have
\begin{align}
&\Bigg\|\prod_{k=s}^{t-1}\bm{P}_{k+1}\mathscr{A}^{\top}\bar{\bm{P}}_{k+1}^{-1}(I_{mn}-\bm{A}_{k})\Bigg\|^{2}\nonumber\\
=&\max_{\|\bm{x}_{s}\|=1}\|\bm{x}_{t}\|^{2}=\max_{\|\bm{x}_{s}\|=1}\|\bm{x}_{t}\bm{P}_{t}^{-1/2}\bm{P}_{t}^{1/2}\|^{2}\nonumber\\
\leq &\max_{\|\bm{x}_{s}\|=1}\|\bm{x}_{t}\bm{P}_{t}^{-1/2}\|^{2}\|\bm{P}_{t}^{1/2}\|^{2}=\max_{\|\bm{x}_{s}\|=1}V_{t}\|\bm{P}_{t}\|\nonumber\\
\leq &\Bigg\{\prod_{k=s}^{t-1}\Bigg(1-\frac{1}{1+\|\bm{Q}_{k}^{-1}\bar{\bm{P}}_{k+1}\|}\Bigg)\Bigg\}\Big\{\|\bm{P}_{t}\|\max_{\|x_{s}\|=1}V_{s}\Big\}\nonumber\\
\leq &\Bigg\{\prod_{k=s}^{t-1}\Bigg(1-\frac{1}{1+\|\bm{Q}_{k}^{-1}\bar{\bm{P}}_{k+1}\|}\Bigg)\Bigg\}\Big\{\|\bm{P}_{t}\|\cdot\|\bm{P}_{s}^{-1}\|\Big\}.
\end{align}

This completes the proof.
\end{IEEEproof}

\subsection{Proof of \emph{Lemma \ref{lem1}}}

To accomplish the proof of \emph{Lemma \ref{lem1}}, we also need the following lemmas. The first three lemmas are all about the properties of $S^{0}$ defined by (\ref{S0}), which can be found in \cite{Guo1994Stability}.

\begin{lem}\cite{Guo1994Stability}\label{lem53}
If two sequences $\alpha_{k}$ and $\beta_{k}$ satisfy $0\leq\alpha_{k}\leq\beta_{k}\leq 1, \forall k\geq 0$, then $\{\alpha_{k}\}\in S^{0}(\lambda)$ implies $\{\beta_{k}\}\in S^{0}(\lambda)$.
\end{lem}

\begin{lem}\cite{Guo1994Stability}\label{lem54}
Let $\{\alpha_{k}\}\in S^{0}(\lambda)$ and $\alpha_{k}\leq\alpha^{*}<1, \forall k\geq 0$ where $\alpha^{*}$ is a constant. Then for any $\epsilon\in(0,1)$, $\{\epsilon\alpha_{k}\}\in S^{0}(\lambda^{(1-\alpha^{*})\epsilon})$.
\end{lem}

\begin{lem}\cite{Guo1994Stability}\label{lem55}
Let $\alpha=\{\alpha_{k},\mathcal{F}_{k}\}$ and $\beta=\{\beta_{k},\mathcal{F}_{k}\}$ be adapted processes, such that
$$
\alpha_{k}\in[0,1],~~~~~\mathbb{E}[\alpha_{k+1}\vert\mathcal{F}_{k}]\geq\beta_{k},~~~~~k\geq 0.
$$
Then $\{\beta\}\in S^{0}(\lambda)$ implies that $\{\alpha\}\in S^{0}(\sqrt{\lambda})$.
\end{lem}

\begin{lem}\label{lem56}
Let $\{\bm{P}_{k}\}$ be generated by (\ref{7}). Then
\begin{equation}
T_{s+1}\leq(1-b_{s+1})T_{s}+d,
\end{equation}
where
\begin{align}
&T_{s}=\sum_{k=(s-1)h^{'}+D_{\mathcal{G}}}^{sh^{'}-1}Tr(\bm{P}_{k+1}),~~~~T_{0}=0,\nonumber
\end{align}
\begin{align}
&b_{s+1}=\frac{a_{min}^{2}c_{s+1}^{1}}{nhc_{s+1}^{2}},\nonumber\\
&c_{s+1}^{1}=Tr\Big(\Big(\sum_{j=1}^{n}P_{sh^{'},j}+h^{'}Q\Big)^{2}\sum_{j=1}^{n}\sum_{sh^{'}+D_{\mathcal{G}}}^{(s+1)h^{'}-1}\frac{\bm{\varphi}_{k,j}\bm{\varphi}_{k,j}^{\top}}{1+\|\bm{\varphi}_{k,j}\|^{2}}\Big),\nonumber\\
&c_{s+1}^{2}=\sum_{j=1}^{n}(r_{j}+1)\cdot\Big(1+\lambda_{max}\Big\{\sum_{j=1}^{n}P_{sh^{'},j}+h^{'}Q\Big\}\Big)\nonumber\\
&~~~~~~~~\cdot Tr\Big(\sum_{j=1}^{n}P_{sh^{'}, j}+h^{'}Q\Big),\nonumber\\
&d=\frac{3}{2}nh(h^{'}+1)Tr(Q),
\end{align}
and $a_{min}=\min\limits_{i,j\in\mathcal{V}}a_{ij}^{(D_{\mathcal{G}})}>0$, $h^{'}=h+D_{\mathcal{G}}$, and $h$ is the constant appearing in \emph{Assumption \ref{asminformation}}.
\end{lem}

The proof of \emph{Lemma \ref{lem56}} is given in the appendix part. The proof of \emph{Lemma \ref{lem1}} is given in the following part:

\begin{IEEEproof}
Denote $\mathcal{H}_{s}=\mathcal{F}_{sh^{'}-1}$. Then it is clear that $T_{s}$ and $b_{s}$  are $\mathcal{H}_{s}$-measurable, and
\begin{align}
b_{s+1}\in\Bigg[0,\frac{a_{min}^{2}}{\sum_{i=1}^{n}(r_{i}+1)}\Bigg],
\end{align}
and
\begin{align}\label{bs}
\mathbb{E}[b_{s+1}\vert\mathcal{H}_{s}]\geq\frac{a_{min}^{2}h^{'}\|Q\|\lambda_{s}^{'}}{m\Big(\sum_{i=1}^{n}r_{i}+n\Big)(1+h^{'}\|Q\|)},
\end{align}
where
$$
\lambda_{s}^{'}=\frac{1}{n(1+h)}\lambda_{min}\Bigg\{\sum_{j=1}^{n}\sum_{sh^{'}+D_{\mathcal{G}}}^{(s+1)h^{'}-1}\frac{\bm{\varphi}_{k,j}\bm{\varphi}_{k,j}^{\top}}{1+\|\bm{\varphi}_{k,j}\|^{2}}\Bigg\}.
$$
By this condition and applying \emph{Lemmas \ref{lem53}, \ref{lem54}} and \emph{\ref{lem55}}, it is easy to see that $\{b_{k+1}\}\in S^{0}(\gamma)$ for some $\gamma\in[0,1)$. Consequently, by the definition of $S^{0}$, we can obtain that
\begin{align}
\mathbb{E}\Bigg[\sum_{k=s}^{t}(1-b_{k+1})\Bigg]\leq C\gamma^{t-s+1}, \forall t\geq s\geq 0,
\end{align}
for some constants $C>0$ and $\gamma\in(0,1)$.

Next, from \emph{Lemma \ref{lem56}}, it follows that for any $\varepsilon>0$
$$
\exp(\varepsilon T_{s+1})\leq \exp((1-b_{s+1})\varepsilon T_{s})\cdot\exp(d\varepsilon).
$$
Consequently, noticing the following inequality
$$
\exp(\alpha x)-1\leq\alpha\exp(x), 0<\alpha<1, x>0,
$$
we get
\begin{equation}
\exp(\varepsilon T_{s+1})\leq \exp(d\varepsilon)\cdot[(1-b_{s+1})\exp(\varepsilon T_{s})+1].
\end{equation}

Hence from this it is easy to know that if $\varepsilon^{*}$ is taken small enough such that $\exp(d\varepsilon)\gamma<1$, then
$$
\sup_{s\geq 0}\mathbb{E}[\exp(\varepsilon T_{s})]<\infty, \forall \varepsilon\in(0,\varepsilon^{*}).
$$
This completes the proof.
\end{IEEEproof}

\subsection{Proof of \emph{Lemma \ref{lem2}}}

Denote
$$
x_{s}=\frac{h(1+\|\bm{Q}_{diag}^{-1}\|\cdot\|\bm{Q}_{diag}\|)+\|\bm{Q}_{diag}^{-1}\|T_{s}}{\mu},
$$
where $T_{s}$ is defined in \emph{Lemma \ref{lem56}}. Then we have
$$
x_{s+1}\leq (1-b_{s+1})x_{s}+\frac{h(1+\|\bm{Q}_{diag}^{-1}\|\cdot\|\bm{Q}_{diag}\|)+d\|\bm{Q}_{diag}^{-1}\|}{\mu}.
$$

It is easy to see from (\ref{bs}), \emph{Assumption \ref{asminformation}} and \emph{Lemma \ref{lem54}} that \emph{Lemma 3.1} in \cite{Guo1994Stability} is applicable to the above equation. Hence we know that
$$
\Bigg\{\frac{1}{x_{s}}\Bigg\}\in S^{0}(\gamma),
$$
for some $\gamma\in(0,1)$. Note that
$$
x_{s}=\sum_{k=(s-1)h^{'}+D_{\mathcal{G}}}^{sh^{'}-1}\frac{1+\|\bm{Q}_{diag}^{-1}\|\cdot\|\bm{Q}_{diag}\|+\|\bm{Q}_{diag}^{-1}\|Tr(\bm{P}_{k+1})}{\mu}.
$$

Hence, similar to the proof in Lemma 5 of \cite{Guo1990Estimating}, it is easy to see that
$$
\Bigg\{\frac{\mu}{1+\|\bm{Q}_{diag}^{-1}\|\cdot\|\bm{Q}_{diag}\|+\|\bm{Q}_{diag}^{-1}\|Tr(\bm{P}_{k})}\Bigg\}\in S^{0}(\lambda),
$$
holds for some $\lambda\in(0,1)$. Then we know that
$$
\Bigg\{\frac{\mu}{1+\|\bm{Q}_{diag}^{-1}\|\cdot\|\bm{Q}_{diag}\|+\|\bm{Q}_{diag}^{-1}\|\cdot\|\bm{P}_{k}\|}\Bigg\}\in S^{0}(\lambda).
$$
Since $(\bar{\bm{P}}_{k+1}-\bm{Q}_{diag})^{-1}=\bm{P}_{k}^{-1}+\bm{R}^{-1}\bm{\Phi}_{k}\bm{\Phi}_{k}$, we have
$$
\bar{\bm{P}}_{k+1}\leq\bm{P}_{k}+\bm{Q}_{diag},
$$
and
$$
\|\bm{Q}_{diag}^{-1}\|\cdot\|\bar{\bm{P}}_{k+1}\|\leq\|\bm{Q}_{diag}^{-1}\|\cdot\|\bm{P}_{k}\|+\|\bm{Q}_{diag}^{-1}\|\cdot\|\bm{Q}_{diag}\|.
$$

By this and \emph{Lemma \ref{lem55}}, we know that
$$
\Bigg\{\frac{\mu}{1+\|\bm{Q}_{diag}^{-1}\|\cdot\|\bar{\bm{P}}_{k}\|}\Bigg\}\in S^{0}(\lambda),
$$
holds for some $\lambda\in(0,1)$.

\section{Simulation Results}

Here we construct a simulation example to illustrate that for regressors that are generated by linear stochastic state space models (where the regressors are strongly correlated and satisfy our cooperative information condition),  even none of the sensors can estimate the parameters individually, the whole sensor network can still fulfill the filtering task cooperatively and effectively. Let us take $n=3$ with the following adjacency matrix
$$
\mathcal{A}=
\begin{pmatrix}
1/3 & 2/3 & 0\\
0 & 1/3 & 2/3 \\
2/3 & 0 & 1/3
\end{pmatrix},
$$
then the corresponding graph is directed, balanced, and strongly connected. 

We will estimate or track an unknown $3$-dimensional signal $\bm{\theta}_{k}$, and assume that the parameter variation $\bm{\delta}_{k} \sim N(0,0.3,3,1)$ (Gaussian distribution) in (\ref{iteration}). In both cases, the observation noises $\{v_{k,i},k\geq 1,i=1,2,3\}$ are independent and identically distributed with $v_{k,i} \sim N(0,0.3,1,1)$ in (\ref{model}), where $\bm{\varphi}_{k,i}(i=1,2,3)$ are generated by a state space model
$$
\begin{cases}
\bm{x}_{k,i}=&A_{i}\bm{x}_{k-1,i}+B_{i}\xi_{k,i},\\
\bm{\varphi}_{k,i}=&C_{i}\bm{x}_{k,i},
\end{cases}
$$
where $\{\xi_{k,i},k\geq 1,i=1,2,3\}$ are independent and identically distributed with $\xi_{k,i}\sim N(0,0.3,1,1)$, and
$$
\begin{aligned}
&A_{1}=A_{2}=
\begin{pmatrix}
1/2 & 0 & 0\\
0 & 1/3 & 0 \\
0 & 0 & 1/5
\end{pmatrix},
A_{3}=
\begin{pmatrix}
4/5 & 0 & 0\\
4/5 & 0 & 0 \\
4/5 & 0 & 0
\end{pmatrix},\\
&B_{1}=(1,0,0)^{T},B_{2}=(1,0,0)^{T},B_{3}=(1,0,0)^{T},\\
&C_{1}=
\begin{pmatrix}
1 & 0 & 0\\
0 & 0 & 0 \\
0 & 0 & 0
\end{pmatrix}, C_{2}=
\begin{pmatrix}
0 & 0 & 0\\
1 & 0 & 0 \\
0 & 0 & 0
\end{pmatrix}, C_{3}=
\begin{pmatrix}
0 & 0 & 0\\
0 & 1 & 0 \\
0 & 0 & 1
\end{pmatrix}.
\end{aligned}
$$
It can be verified that \emph{Assumption \ref{asminformation}} is satisfied.

For numerical simulations, let $\bm{x}_{0,1}=\bm{x}_{0,2}=\bm{x}_{0,3}=(1,1,1)^{\top}, \bm{\theta}_{0}=(1,1,1)^{\top},\hat{\bm{\theta}}_{0,i}=(0,0,0)^{\top}, P_{0,i}=I_{3}, r_{i}=0.1  (i=1,2,3)$ and $Q=0.1\times I_{3}$. Here we repeat the simulation for $m=500$ times with the same initial states. Then for sensor $i(i=1,2,3)$, we can get $m$ sequences 
$$
\{\|\hat{\bm{\theta}}_{k,i}^{j}-\bm{\theta}^{j}_{k}\|^{2}, k=1, 100, 200, \dots, 2000\},~~~~j=1, \dots, m,
$$ 
where the superscript $j$ denotes the $j$-th simulation result. We use 
$$
\frac{1}{m}\sum_{j=1}^{m}\|\hat{\bm{\theta}}_{k,i}^{j}-\bm{\theta}_{k}^{j}\|^{2},~~~~k=1, 100, 200, \dots, 2000,
$$ 
to approximate the tracking errors in Fig. 1.

\begin{figure}[!htb]
\begin{center}
\renewcommand{\captionfont}{\footnotesize}
 \includegraphics[width=\hsize]{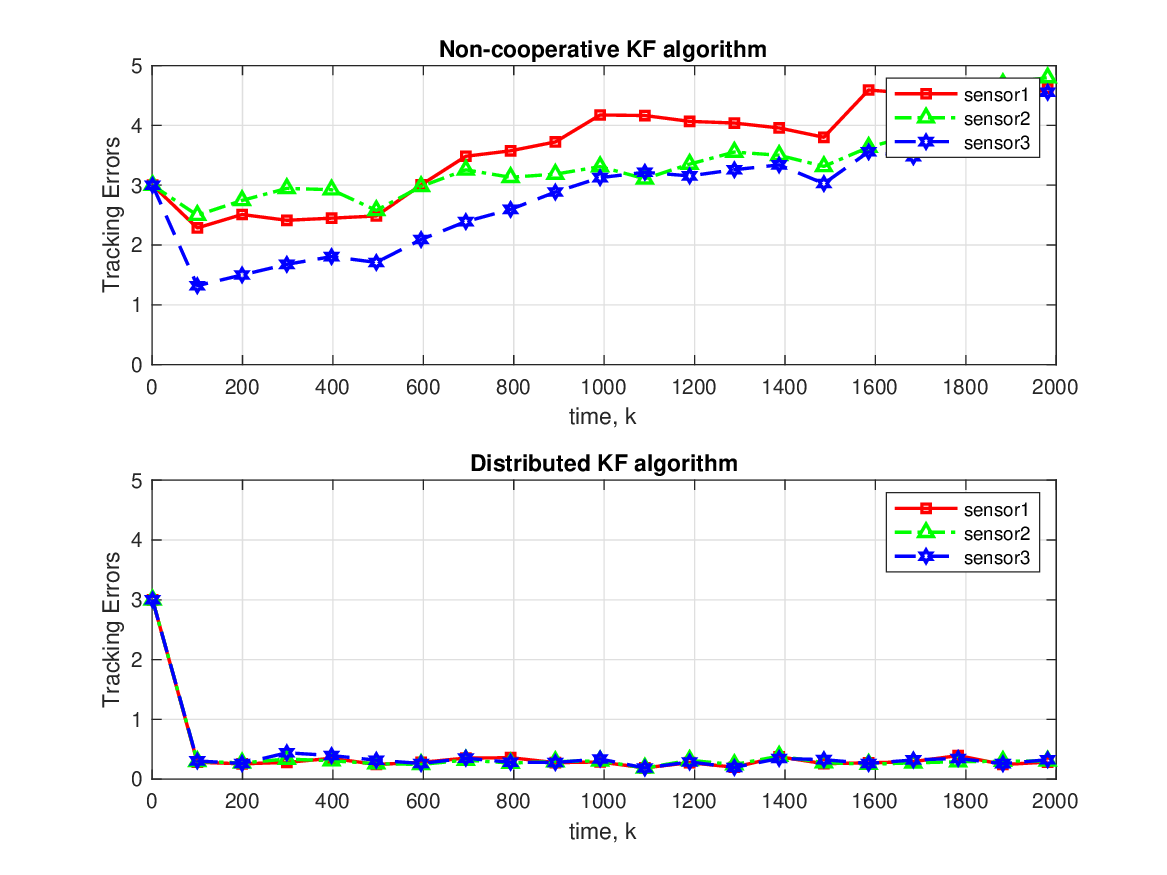}
 \caption{Tracking errors of the three sensors}
\end{center}
\end{figure}

The upper one in Fig. 1 is the non-cooperative KF algorithm in which the tracking errors of the three sensors are all quite large because all the sensors do not satisfy the information condition in \cite{Guo1994Stability}. The lower one in Fig. 1 is the distributed KF algorithm in which all the tracking errors converge to a small neighborhood of zero as $k$ increases, since the whole system satisfies \emph{Assumption \ref{asminformation}}. 

\section{Concluding Remarks}

In this paper, we have provided the stability analysis for a distributed KF algorithm, which can be used to track a time-varying parameter vector cooperatively in sensor networks. Here we need no independence, no stationarity and no Gaussian property for the stability analysis, which makes it possible for our theory to be applicable to stochastic feedback systems, and lays a foundation for further investigation on related problems concerning the combination of learning, communication and control. In addition, the cooperative excitation condition used in the paper implies that the distributed KF algorithm can accomplish the tracking task cooperatively, even if any individual sensor cannot due to lack of necessary excitation. Of course, there are still a number of interesting problems for further research, for examples, to consider more general system models with random coefficients, and to combine distributed adaptive filters with distributed control problems, etc.

\begin{appendices}

\section{Some basic lemmas}

\begin{lem}\cite{Guo1993Time}\label{lemA1}
For any matrices $A, B, C$ and $D$ with suitable dimensions,
$$
(A+BCD)^{-1}=A^{-1}-A^{-1}B(D^{-1}+CA^{-1}B)^{-1}CA^{-1},
$$
holds, provided that the relevant matrices are invertible.
\end{lem}

\begin{lem}\cite{Guo1993Time}\label{lemA2}
Let $A\in\mathbb{R}^{d\times s}$ and  $B\in\mathbb{R}^{s\times d}$ be two matrices. Then the nonzero eigenvalues of the matrices $AB$ and $BA$ are the same, and $\vert I_{d}+AB\vert=\vert I_{s}+BA\vert$ holds. Moreover, if $d=s$, then $\vert AB\vert=\vert A\vert\cdot\vert B\vert=\vert BA\vert, \text{Tr}(A)=\text{Tr}(A^{\top}),\text{Tr}(AB)=\text{Tr}(BA)$. Furthermore, if $A$ and $B$ are positive definite matrices with $A\geq B$, then $A^{-1}\leq B^{-1}$.
\end{lem}

\begin{lem}\cite{Guo1993Time}\label{lemA3}
For any scalar sequences $a_{j}\geq 0, b_{j}\geq 0, (j=1,\dots,m)$, the following inequalities hold:
\begin{itemize}
\item $C_{r}$-inequality:
$$
\Bigg(\sum\limits_{j=1}^{m}a_{j}\Bigg)^{r}\leq
\begin{cases}
&m^{r-1}\sum\limits_{j=1}^{m}a_{j}^{r},~~~~r\geq 1,\\
&\sum\limits_{j=1}^{m}a_{j}^{r},~~~~~~~~~~0\leq r\leq 1.
\end{cases}
$$

\item Schwarz inequality:
$$
\sum_{j=1}^{n}a_{j}b_{j}\leq \Bigg\{\sum_{j=1}^{n}a_{j}^{2}\Bigg\}^{\frac{1}{2}}\Bigg\{\sum_{j=1}^{n}b_{j}^{2}\Bigg\}^{\frac{1}{2}}.
$$
\end{itemize}
\end{lem}

\begin{rem}
By $C_{r}$- and Schwarz inequalities, it is easy to obtain that
$$
\sum_{j=1}^{n}a_{j}b_{j}\leq \sum_{j=1}^{n}a_{j}\sum_{j=1}^{n}b_{j}.
$$
Furthermore, by choosing $a_{j}=\frac{c_{j}}{d_{j}}, b_{j}=d_{j}$, where $c_{j}\geq 0, d_{j}>0$, then it easy to conclude that
$$
\sum_{j=1}^{n}\frac{c_{j}}{d_{j}}\geq\frac{\sum_{j=1}^{n}c_{j}}{\sum_{j=1}^{n}d_{j}}.
$$
\end{rem}

\section{Proof of \emph{Lemma \ref{lem56}}}

For ease of representation, we let $a_{ij}^{(s)}$ be the $(i,j)$th entry of the matrix $\mathcal{A}^{s}, s\geq 1$, where $a_{ij}^{(1)}=a_{ij}$. By (\ref{7}), we have
\begin{align}
P_{k,i}=&\Bigg\{\sum_{j=1}^{n}a_{ji}\bar{P}_{k,j}^{-1}\Bigg\}^{-1}\leq\sum_{j=1}^{n}a_{ji}\bar{P}_{k,j}\nonumber\\
=&\sum_{j=1}^{n}a_{ji}(\bar{P}_{k,j}-Q)+Q\nonumber\\
=&\sum_{j=1}^{n}a_{ji}(P_{k-1,j}^{-1}+r_{j}^{-1}\bm{\varphi}_{k-1,j}\bm{\varphi}_{k-1,j}^{\top})^{-1}+Q\nonumber\\
\leq&\sum_{j=1}^{n}a_{ji}P_{k-1,j}+Q\nonumber\\
\leq & \sum_{j=1}^{n}a_{ji}\Bigg(\sum_{t=1}^{n}a_{tj}P_{k-2,t}\Bigg)+2Q\nonumber\\
=&\sum_{j=1}^{n}a_{ji}^{(2)}P_{k-2,j}+2Q\nonumber\\
\leq&\dots\nonumber\\
\leq & \sum_{j=1}^{n}a_{ji}^{(k-sh^{'})}P_{sh^{'},j}+(k-sh^{'})Q\nonumber\\
\leq & \sum_{j=1}^{n}a_{ji}^{(k-sh^{'})}P_{sh^{'},j}+h^{'}Q,
\end{align}
holds for any $k\in[sh^{'}+D_{\mathcal{G}},(s+1)h^{'}]$. Hence by the matrix inverse formula, i.e., \emph{Lemma \ref{lemA1}} in the Appendix, it follows that for any $k\in[sh^{'}+D_{\mathcal{G}},(s+1)h^{'}]$,
\begin{align}
&P_{k+1,i}\nonumber\\
=&\Bigg\{\sum_{j=1}^{n}a_{ji}\bar{P}_{k+1,j}^{-1}\Bigg\}^{-1}\nonumber\\
= & \Bigg\{\sum_{j=1}^{n}a_{ji}[(P_{k,j}^{-1}+r_{j}^{-1}\bm{\varphi}_{k,j}\bm{\varphi}_{k,j}^{\top})^{-1}+Q]^{-1}\Bigg\}^{-1}\nonumber\\
\leq &\sum_{j=1}^{n}a_{ji}(P_{k,j}^{-1}+r_{j}^{-1}\bm{\varphi}_{k,j}\bm{\varphi}_{k,j}^{\top})^{-1}+Q\nonumber\\
\leq & \sum_{j=1}^{n}a_{ji}\Bigg[\Bigg(\sum_{t=1}^{n}a_{tj}^{(k-sh^{'})}P_{sh^{'},t}+h^{'}Q\Bigg)^{-1}+r_{j}^{-1}\bm{\varphi}_{k,j}\bm{\varphi}_{k,j}^{\top}\Bigg]^{-1}\nonumber\\
&+Q\nonumber
\end{align}
\begin{align}
= & Q+\sum_{j=1}^{n}a_{ji}\Bigg[\sum_{t=1}^{n}a_{tj}^{(k-sh^{'})}P_{sh^{'},t}+h^{'}Q\nonumber\\
&-\Big(\sum_{t=1}^{n}a_{tj}^{(k-sh^{'})}P_{sh^{'},t}+h^{'}Q\Big)\bm{\varphi}_{k,j}\bm{\varphi}_{k,j}^{\top}\nonumber\\
&\cdot \frac{\Big(\sum\limits_{t=1}^{n}a_{tj}^{(k-sh^{'})}P_{sh^{'},t}+h^{'}Q\Big)}{r_{j}+\bm{\varphi}_{k,j}^{\top}\Big(\sum\limits_{t=1}^{n}a_{tj}^{(k-sh^{'})}P_{sh^{'},t}+h^{'}Q\Big)\bm{\varphi}_{k,j}}\Bigg]\nonumber\\
= & \sum_{j=1}^{n}a_{ji}^{(k-sh^{'}+1)}P_{sh^{'},j}+(h^{'}+1)Q\nonumber\\
&-\sum_{j=1}^{n}a_{ji}\Big(\sum_{t=1}^{n}a_{tj}^{(k-sh^{'})}P_{sh^{'},t}+h^{'}Q\Big)\bm{\varphi}_{k,j}\bm{\varphi}_{k,j}^{\top}\nonumber\\
&\cdot \frac{\Big(\sum\limits_{t=1}^{n}a_{tj}^{(k-sh^{'})}P_{sh^{'},t}+h^{'}Q\Big)}{r_{j}+\bm{\varphi}_{k,j}^{\top}\Big(\sum\limits_{t=1}^{n}a_{tj}^{(k-sh^{'})}P_{sh^{'},t}+h^{'}Q\Big)\bm{\varphi}_{k,j}}\nonumber\\
\leq  &\sum_{j=1}^{n}a_{ji}^{(k-sh^{'}+1)}P_{sh^{'},j}+(h^{'}+1)Q\nonumber\\
&-\sum_{j=1}^{n}a_{ji}\Big(\sum\limits_{t=1}^{n}a_{tj}^{(k-sh^{'})}P_{sh^{'},t}+h^{'}Q\Big)\frac{\bm{\varphi}_{k,j}\bm{\varphi}_{k,j}^{\top}}{1+\|\bm{\varphi}_{k,j}\|^{2}}\nonumber\\
&\cdot \frac{\Big(\sum\limits_{t=1}^{n}a_{tj}^{(k-sh^{'})}P_{sh^{'},t}+h^{'}Q\Big)}{(r_{j}+1)\Big(1+\lambda_{max}\Big\{\sum\limits_{t=1}^{n}a_{tj}^{(k-sh^{'})}P_{sh^{'},t}+h^{'}Q\Big\}\Big)}.
\end{align}

By \emph{Assumption \ref{asminformation}} and \emph{Remark 3.1}, we know that $a_{ji}^{(D_{\mathcal{G}})}\geq a_{min}>0$, where $a_{min}=\min\limits_{i,j\in\mathcal{V}}a_{ij}^{(D_{\mathcal{G}})}>0$, where $D_{\mathcal{G}}$ is the diameter of the graph $\mathcal{G}$. Consequently, it is not difficult to see that for any $k>D_{\mathcal{G}}$,  $a_{ji}^{(k)}\geq a_{min}$ holds. Then for $k\in[sh^{'}+D_{\mathcal{G}},(s+1)h^{'}]$, we have by noting inequalities in \emph{Remark A.1} that
\begin{align}
&Tr(\bm{P}_{k+1})\nonumber\\
=&Tr\Bigg(\sum_{i=1}^{n}P_{k+1,i}\Bigg)\nonumber\\
\leq  &Tr\Bigg(\sum_{i=1}^{n}\sum_{j=1}^{n}a_{ji}^{(k-sh^{'}+1)}P_{sh^{'},j}\Bigg)+n(h^{'}+1)Tr(Q)\nonumber\\
&-Tr\Bigg(\sum_{i=1}^{n}\sum_{j=1}^{n}a_{ji}\Big[\sum_{t=1}^{n}a_{tj}^{(k-sh^{'})}P_{sh^{'},t}+h^{'}Q\Big]\frac{\bm{\varphi}_{k,j}\bm{\varphi}_{k,j}^{\top}}{1+\|\bm{\varphi}_{k,j}\|^{2}}\nonumber\\
&\cdot \frac{\sum\limits_{t=1}^{n}a_{tj}^{(k-sh^{'})}P_{sh^{'},t}+h^{'}Q}{(r_{j}+1)\Big(1+\lambda_{max}\Big\{\sum\limits_{t=1}^{n}a_{tj}^{(k-sh^{'})}P_{sh^{'},t}+h^{'}Q\Big\}\Big)}\Bigg)\nonumber\\
=  &Tr\Bigg(\sum_{j=1}^{n}P_{sh^{'},j}\Bigg)+n(h^{'}+1)Tr(Q)\nonumber\\
&-\sum_{j=1}^{n}\frac{1}{(r_{j}+1)\Big(1+\lambda_{max}\Big\{\sum\limits_{t=1}^{n}a_{tj}^{(k-sh^{'})}P_{sh^{'},t}+h^{'}Q\Big\}\Big)}\nonumber
\end{align}
\begin{align}
&\cdot Tr\Big(\Big[\sum_{t=1}^{n}a_{tj}^{(k-sh^{'})}P_{sh^{'},t}+h^{'}Q\Big]\frac{\bm{\varphi}_{k,j}\bm{\varphi}_{k,j}^{\top}}{1+\|\bm{\varphi}_{k,j}\|^{2}}\nonumber\\
&\cdot\Big[\sum_{t=1}^{n}a_{tj}^{(k-sh^{'})}P_{sh^{'},t}+h^{'}Q\Big]\Big)\nonumber\\
\leq  &Tr(\bm{P}_{sh^{'}})+n(h^{'}+1)Tr(Q)\nonumber\\
&-\frac{1}{\sum\limits_{j=1}^{n}(r_{j}+1)\Big(1+\lambda_{max}\Big\{\sum\limits_{t=1}^{n}a_{tj}^{(k-sh^{'})}P_{sh^{'},t}+h^{'}Q\Big\}\Big)}\nonumber\\
&\cdot Tr\Big(\sum_{j=1}^{n}\Big[\sum_{t=1}^{n}a_{tj}^{(k-sh^{'})}P_{sh^{'},t}+h^{'}Q\Big]\frac{\bm{\varphi}_{k,j}\bm{\varphi}_{k,j}^{\top}}{1+\|\bm{\varphi}_{k,j}\|^{2}}\nonumber\\
&\cdot\Big[\sum_{t=1}^{n}a_{tj}^{(k-sh^{'})}P_{sh^{'},t}+h^{'}Q\Big]\Big)\nonumber\\
\leq  &Tr(\bm{P}_{sh^{'}})+n(h^{'}+1)Tr(Q)\nonumber\\
&-\frac{Tr\Bigg(\sum\limits_{j=1}^{n}P_{sh^{'},j}\Bigg)}{\sum\limits_{j=1}^{n}(r_{j}+1)\cdot\sum\limits_{j=1}^{n}\Big(1+\lambda_{max}\Big\{\sum\limits_{t=1}^{n}a_{tj}^{(k-sh^{'})}P_{sh^{'},t}+h^{'}Q\Big\}\Big)}\nonumber\\
&\cdot \frac{1}{Tr\Big(\sum\limits_{j=1}^{n}P_{sh^{'},j}+h^{'}Q\Big)}\nonumber\\
&\cdot Tr\Big(\sum_{j=1}^{n}\Big[\sum_{t=1}^{n}a_{tj}^{(k-sh^{'})}P_{sh^{'},t}+h^{'}Q\Big]\frac{\bm{\varphi}_{k,j}\bm{\varphi}_{k,j}^{\top}}{1+\|\bm{\varphi}_{k,j}\|^{2}}\nonumber\\
&\cdot \Big[\sum_{t=1}^{n}a_{tj}^{(k-sh^{'})}P_{sh^{'},t}+h^{'}Q\Big]\Big)\nonumber\\
\leq  &Tr(\bm{P}_{sh^{'}})+n(h^{'}+1)Tr(Q)\nonumber\\
&-\frac{Tr(\bm{P}_{sh^{'}})}{n\sum\limits_{j=1}^{n}(r_{j}+1)\cdot\Big(1+\lambda_{max}\Big\{\sum\limits_{t=1}^{n}P_{sh^{'},t}+h^{'}Q\Big\}\Big)}\nonumber\\
&\cdot \frac{a_{min}^{2}Tr\Big(\Big(\sum\limits_{t=1}^{n}P_{sh^{'},t}+h^{'}Q\Big)^{2}\sum\limits_{j=1}^{n}\frac{\bm{\varphi}_{k,j}\bm{\varphi}_{k,j}^{\top}}{1+\|\bm{\varphi}_{k,j}\|^{2}}\Big)}{Tr\Big(\sum\limits_{j=1}^{n}P_{sh^{'},j}+h^{'}Q\Big)}.
\end{align}

Summing both sides, we obtain
\begin{align}
&T_{s+1}\nonumber\\
=&\sum_{k=sh^{'}+D_{\mathcal{G}}}^{(s+1)h^{'}-1}Tr(\bm{P}_{k+1})\nonumber\\
\leq & hTr(\bm{P}_{sh})+nh(h^{'}+1)Tr(Q)\nonumber\\
&-\frac{a_{min}^{2}hTr(\bm{P}_{sh^{'}})}{nh\sum\limits_{j=1}^{n}(r_{j}+1)\cdot\Big(1+\lambda_{max}\Big\{\sum\limits_{t=1}^{n}P_{sh^{'},t}+h^{'}Q\Big\}\Big)}\nonumber\\
&\cdot \frac{Tr\Big(\Big(\sum\limits_{t=1}^{n}P_{sh^{'},t}+h^{'}Q\Big)^{2}\sum\limits_{j=1}^{n}\sum\limits_{k=sh^{'}+D_{\mathcal{G}}}^{(s+1)h^{'}-1}\frac{\bm{\varphi}_{k,j}\bm{\varphi}_{k,j}^{\top}}{1+\|\bm{\varphi}_{k,j}\|^{2}}\Big)}{Tr\Big(\sum\limits_{j=1}^{n}P_{sh^{'},j}+h^{'}Q\Big)}\nonumber\\
\leq & hTr(\bm{P}_{sh^{'}})-b_{s+1}hTr(\bm{P}_{sh^{'}})+nh(h^{'}+1)Tr(Q).
\end{align}

Again we have
\begin{align}
&hTr(\bm{P}_{sh^{'}})\nonumber\\
=&\sum_{k=(s-1)h^{'}+D_{\mathcal{G}}}^{sh^{'}-1}\sum_{j=1}^{n}Tr(P_{sh^{'},j})\nonumber\\
\leq &\sum_{k=(s-1)h^{'}+D_{\mathcal{G}}}^{sh^{'}-1}\sum_{j=1}^{n}Tr\Bigg(\sum_{t=1}^{n}a_{t,j}^{(sh^{'}-k)}P_{k+1,t}+(sh^{'}-k)Q\Bigg)\nonumber\\
=&T_{s}+\frac{1}{2}nh(h^{'}+1)Tr(Q),
\end{align}
and
\begin{align}
T_{s+1}\leq &(1-b_{s+1})T_{s}+\frac{3}{2}nh(h^{'}+1)Tr(Q)\nonumber\\
=&(1-b_{s+1})T_{s}+d, s\geq 0.
\end{align}
This completes the proof.
\end{appendices}

\bibliographystyle{IEEEtran}
\bibliography{references}

\end{document}